\documentclass[10pt]{IEEEtran}
\usepackage{geometry}                
\usepackage{graphicx}
\usepackage{amssymb,amsmath}
\usepackage{epstopdf}
\usepackage{color} 
\usepackage{hyperref}

\newcommand{\bit}{\begin{itemize}}
\newcommand{\eit}{\end{itemize}}
\newcommand{\bd}{\begin{description}}
\newcommand{\ed}{\end{description}}

\newcommand{\bc}{\begin{center}}
\newcommand{\ec}{\end{center}}

\newcommand{\be}{\begin{equation}}
\newcommand{\ee}{\end{equation}}
\newcommand{\bea}{\begin{eqnarray}}
\newcommand{\eea}{\end{eqnarray}}
\newcommand{\bs}{\begin{subequations}}
\newcommand{\es}{\end{subequations}}
\newcommand{\nn}{\nonumber}

\newcommand{\bra}[1]{\langle {#1}|}
\newcommand{\ket}[1]{|{#1}\rangle}
\newcommand{\braket}[2]{\langle {#1} | {#2}\rangle}

\newcommand{\lightline}{\rule{\textwidth}{0.2pt}}

\DeclareMathOperator{\Tr}{Tr}

\title{Optimal universal learning machines for quantum state discrimination}
\author{Marco Fanizza, Andrea Mari, Vittorio Giovannetti
\thanks{Marco Fanizza is with NEST, Scuola Normale Superiore, Istituto Nanoscienze-CNR, I-56126 Pisa, Italy.}
\thanks{Andrea Mari is with NEST, Scuola Normale Superiore, Istituto Nanoscienze-CNR, I-56126 Pisa, Italy.
Now at Xanadu, 777 Bay Street, Toronto, Ontario, M5G 1S5, Canada.
}
\thanks{Vittorio Giovannetti is with NEST, Scuola Normale Superiore and Istituto Nanoscienze-CNR, I-56126 Pisa, Italy.}}

\begin{document}
\maketitle
\begin{abstract}
We consider the problem of correctly classifying a given quantum two-level system (qubit) which is known to be in one of two equally probable quantum states. We assume that this task should be performed by a quantum machine which does not have at its disposal a complete classical description of the two template states, but can only have partial prior information about their level of purity and mutual overlap. Moreover, similarly to the classical supervised learning paradigm, we assume that the machine can be trained by $n$ qubits prepared in the first template state and by $n$ more qubits prepared in the second template state. In this situation we are interested in the optimal process which correctly classifies the input qubit with the largest probability allowed by quantum mechanics. The problem is studied in its full generality for a number of different prior information scenarios and for an arbitrary size $n$ of the training data. Finite size corrections around the asymptotic limit $n\rightarrow \infty$ are derived. When the states are assumed to be pure, with known overlap, the problem is also solved in the case of d-level systems.
\end{abstract}

\section{Introduction}

Machine Learning (ML) is that branch of computer science which studies how to instruct a computer to solve a specific task
 by feeding it with 
 a collection of training data from which it could learn how to proceed. This approach finds applications  in a variety of practical  pattern recognition, decision and clustering problems where  a definite classification of the various alternatives are not directly accessible~\cite{MACHINE}. 
Not surprisingly, the interplay between ML and quantum information is very promising (see   \cite{WITTEK}, \cite{Schuld}, \cite{BRIEGEL}, \cite{natureqml}, \cite{Cilibertoqml} and references therein). ML has been proposed as a useful tool to improve the performances of a variety of quantum information procedures, 
e.g.  identification of optimal quantum measurement and estimation procedures, quantum gate design and quantum dynamics engineering. On the other hand, it has been shown that quantum computing can provide speed-ups for ML problems. 
Moreover, as originally hinted in Refs.~\cite{SCJ,SC,HHH1,HHH2,Gambs},
a drastic departure from classical data analysis is instead realized in Quantum Learning (QL), where the ``learning from examples" paradigm  is adopted as a new mode of operation 
of quantum devices which have access to (not necessarily classical) training data. 
This setting 
appears to be perfectly suited to deal with 
 the specific character of quantum mechanics where, at variance with 
classical models, 
a fundamental discrepancy exists between the state of a system and the ``knowledge" one can acquire about it through measurements. Such discrepancy is a distintive feature of the theory:  ultimately it can be traced back to the no-cloning theorem~\cite{NoCloning} and poses intrinsic limitations on information retrieval processes.
Accordingly in quantum mechanics, the ability of perfectly discriminating alternative configurations, let them being states or processes, can only be guaranteed under special conditions (semiclassical limit).
Since the seminal works of Helstrom~\cite{Hel}, Holevo~\cite{HolOp} and Yuen \textit{et al.}~\cite{YKL}, developing optimal probabilistic strategies to face  these limitations is a fundamental problem of quantum information. 
A standard example is provided by quantum state discrimination: here an agent is presented with a quantum system $Q$ and asked to
identify  its  state knowing that the latter was randomly drawn from an ensemble of possible alternatives 
which are 
  specified in terms of classical data  that fully characterize them. 
 The QL version of this problem is obtained by replacing such  classical descriptions with a collection of
 quantum ancillary systems initialized into the same template states 
 the agent has to assign to $Q$.
A universal machine for optimal discrimination is hence identified as the quantum device 
which, by having full access to $Q$ and  the ancillas, allows the agent to solve the identification task 
with the smallest probability of error. 
The problem has been addressed in various scenarios in~\cite{HHH1,HHH2,BH,GK,Sentisprogram,sentismemory,SGA,Sentismargin}, and has attracted the attention of the community as an example of a genuine supervised QL task ~\cite{BRIEGEL,natureqml,inductiveqml}.
In this article we present results about universal machines for qubit discrimination, which can discriminate among any two states, extending in particular the results of~\cite{Sentisprogram} to include a variety of scenarios. 
Specifically we focus on hybrid QL configurations where the agent, beside  being provided with the quantum ancillas, has also access to some
 prior classical information on the templates configurations, such as their purity or their mutual  distance. 
 These scenarios naturally emerge when, for instance, the training and the target data are effected by some deteriorating processes (say dephasing or decoherence transformations) which 
 the agent cannot prevent from occurring, but whose 
 operating mechanisms are known to him,  or when the different templates are affected by  uncertainties  arising from the absence of a common, shared reference frame.

 The manuscript is organized as follows: 
 notation and  model  are introduced  in Sec.~\ref{SEC:model}. The principal part of the paper  is Sec.~\ref{SEC:DER} where our results are explicitly derived in three dedicated subsections.
In the first two paragraphs we extend the results of the work of Ref.~\cite{Sentisprogram}.
  Specifically in Sec.~\ref{SEC:i} we study the case of an optimal universal machines which is trained to discriminate between two qubit density matrices of fixed but different purities.
 In Sec.~\ref{SEC:ii} instead we focus on the case where the training data are two generic (possibly) mixed quantum systems. In Sec.~\ref{SEC:iii} we discuss the scenario where
 the training data are pure with fixed relative overlap, but otherwise unknown. The interest in this last configuration arises when considering QML processes  where, in analogy with the schemes analyzed in Refs.~\cite{REF1,REF2,REF3,REF4,REF5}, the party who is creating the template states 
does not share a common reference frame with the party that is supposed to solve the identification problem. For this special setting results are extended beyond the
qubit case  to include  arbitrary $d$-dimensional input systems. Finally in Sec.~\ref{SUBSECTIOND} we compare the optimal machines that leads to the optimal probability
thresholds for the three scenarios,  commenting about their compatibility.
Section~\ref{SEC:povm} presents an implementation of optimal machines obtained by exploiting the QISKit software development kit~\cite{qiskit}.

The paper ends with conclusions in Sec.~\ref{SEC:con}. Technical material is presented in dedicated appendices.

\section{The model} \label{SEC:model} 

In a classical supervised learning classification problem an algorithm receives as input a training set of labelled data, and outputs a classifier which can be used to predict the label of new unlabelled data. In a probabilistic setting one can suppose that the dataset, made of couples $(x,y)$ of data $x\in X$ and labels $y\in Y$ obeys a probability distribution $P: X\times Y \rightarrow [0, 1]$. Then, a classifier is a labelling rule obeying another probability distribution $C : X \rightarrow Y$. For each $P$ there exists an optimal classifier which minimises the probability of misclassification; a good learning algorithm should obtain a classifier that with a misclassification probability close to the optimal, as the training dataset becomes large, and with the fewest assumptions on the distribution $P$. The assumptions on $P$ can also be described probabilistically as a prior probability distribution $G$ over the possible $P$. Given this prior $G$, one can say that an algorithm is optimal if it attains the lowest probability on average, where the average is done over all the possible distributions $P$, assuming they are distributed according to $G$. A straightforward way to generalise classical probabilistic task is to substitute probability distributions with quantum states: in the problem considered in this article, we substitute conditional probabilities $P(X|Y)$ with quantum states distributed according a classical prior.
More precisely in the scenario we have in mind a qubit system $X$ is initialized with probability $1/2$ in one of two possible template states $\rho_1$ and $\rho_2$.
Without having access to the full classical description of these templates configurations (i.e. without knowing the explicit values of their associated Bloch vectors $\mathbf r_1$ and $\mathbf r_2$, see below), 
an external agent is now asked to  identify which of the two alternative
actually occurred  by only granting him access to $X$ and to two independent  sets of $n$ ancillary qubits $A$ and $B$, initialized respectively into 
$n$ copies of $\rho_1$ and $\rho_2$.
Following Refs.~\cite{HHH1,HHH2,Sentisprogram,sentismemory} 
solutions to this problem can be assigned in terms of  
a two-outcome POVM $\hat{\mathcal M}\equiv\{\hat \Pi_1,\hat \Pi_2\}$ that acts globally on the full system $AXB$ formed by the test qubit  $X$ and by the two $n$-qubits ancillas. 
In particular, noticing that  the possible states of $AXB$ are the density matrices  $\tau_1=  \rho_1^{\otimes n}\otimes\rho_1\otimes \rho_2^{\otimes n}$ (corresponding
to have $X$ in $\rho_1$) and 
$\tau_2= \rho_1^{\otimes n}\otimes\rho_2\otimes \rho_2^{\otimes n}$ ($X$ in $\rho_2$),  the average  error probability of the procedure can be computed
as 
\bea
P_{err}^{(n)}= \int d\mu(\rho_1,\rho_2)  \frac{   \mbox{Tr}[ \tau_1 \hat{\Pi}_2 ] + \mbox{Tr}[ \tau_2 \hat{\Pi}_1 ] }{2},
\eea
 where  $ \mbox{Tr}[ \tau_j \hat{\Pi}_j ]$ is the probability of success in the identifying the $j$-th configuration, while
 $d\mu(\rho_1,\rho_2)$ is a probability measure that gauges  the initial ignorance of the agent about $\rho_1$ and $\rho_2$. 
  Exploiting then the  completeness relation of $\hat{\mathcal M}$ this can be finally recast into 
\begin{eqnarray} \label{QUESTA}
P_{err}^{(n)}  =  \frac{1}{2} - \frac{1}{4} \mbox{Tr}[ \Theta(\hat{\Pi}_1-\hat{\Pi}_2) ] \;, \end{eqnarray} 
where  $\Theta$ is the trace-null, Hermitian operator
\bea \label{DEFo} 
\Theta= {\alpha^{(n)}}-{\beta^{(n)}}\;,\eea 
 given by the difference between the following density matrices of $AXB$, 
\bea 
&{\alpha^{(n)}}&\equiv \int d\mu(\rho_1,\rho_2) \rho_1^{\otimes n}\otimes\rho_1\otimes \rho_2^{\otimes n},\nn\\& {\beta^{(n)}}& \equiv\int d\mu(\rho_1,\rho_2) \rho_1^{\otimes n}\otimes\rho_2\otimes \rho_2^{\otimes n}\label{DEFalpha}. 
\eea
By the Holevo-Helstrom theorem \cite{Hel}, the minimum~(\ref{QUESTA}) can now be easily obtained by choosing an optimal POVM $\hat{\mathcal M}$ which has components $\hat \Pi_1$, $\hat \Pi_2$  respectively projecting on the positive and the negative eigenspaces of $\Theta$, i.e. 
\begin{eqnarray} \label{QUESTAqui0}
P_{err,min}^{(n)}  =  \frac{1}{2} - \frac{1}{4} \|\Theta\|_1 \;, \end{eqnarray} 
with the symbol $\|\cdots\|_1$ indicating the  trace norm. 

Some general properties of $P_{err,min}^{(n)}$  can be determined 
by simple arguments. 
First of all since the agent can always discard part of the ancillary states before attempting to identify $Q$, for all possible choices of the measure $d\mu(\rho_1,\rho_2)$,  $P_{err,min}^{(n)}$
has to fulfil the inequality 
\begin{eqnarray} \label{QUESTAqui12}
P_{err,min}^{(n)}  \leq   \frac{1}{2} - \frac{1}{4}   \| \int d\mu(\rho_1,\rho_2) (\rho_1-\rho_2) \|_1\;, \end{eqnarray}
and being a decreasing function of $n$, i.e. 
\begin{eqnarray} P_{err,min}^{(n)} \geq P_{err,min}^{(n+1)}\;.
\end{eqnarray} 
Furthermore, by exploting the joint-convexity of the trace-norm~\cite{HOLEVOBOOK}  the following lower bound can be established 
\begin{eqnarray} \label{QUESTAqui1}
P_{err,min}^{(n)}  \geq  \frac{1}{2} - \frac{1}{4}  \int d\mu(\rho_1,\rho_2)  \|\rho_1-\rho_2\|_1\;,  \end{eqnarray} 
for all $n$ integers. The term on the right-hand-side of this inequality corresponds to the average Helstrom error probability $\bar{P}_H$, i.e. 
the average minimum error probability the agent could attain by providing him/her 
 with a full classical description of the template states: under this condition in fact, for each couple of density matrices $\rho_1$ and $\rho_2$,  he/she can taylor a specific POVM  on $X$  that it is optimized  to  distinguish  them.  
Invoking a full tomographic reconstruction of $\rho_1$ and $\rho_2$, the gap between $P_{err,min}^{(n)}$ and $\bar{P}_H$ (optimal excess risk function~\cite{Sentisprogram,sentismemory}), can be 
 shown to nullify  in the asymptotic regime $n\rightarrow \infty$, i.e.   \begin{equation} \label{QUESTAqui}
\lim_{n\rightarrow \infty} P_{err,min}^{(n)}  =  \frac{1}{2} - \frac{1}{4}  \int d\mu(\rho_1,\rho_2)  \|\rho_1-\rho_2\|_1  \end{equation}

Apart from the above results  explicit expressions for $P_{err,min}^{(n)}$ are known only for a limited set of configurations. 
For instance in Ref. \cite{HHH1} the Authors focus
on the case where both $\rho_1$ and $\rho_2$ are pure in general finite dimension, while Ref. \cite{Sentisprogram} provides the formal solution under the assumption that $\rho_1$ and $\rho_2$ are density matrices having  the same assigned purity.
The aim of the present work is to extend these results by expanding the set of treatable scenarios to include the following cases
\begin{itemize}
\item[i)]$\rho_1$ and $\rho_2$ having different assigned purities but being otherwise arbitrary;  
 \item[ii)]   $\rho_1$ and $\rho_2$ being completely arbitrary (not necessarily pure) density matrices; 
 \item[iii)] $\rho_1$ and $\rho_2$ being arbitrary pure  states having assigned mutual distance.
\end{itemize} 
For these configurations we compute the associated values of $P_{err,min}^{(n)}$ reporting closed analytical expressions for the higher order contributions of their asymptotic  expansions  at large $n$.

\section{Derivation} \label{SEC:DER} 
The key ingredient for deriving the above results is the evaluation of the eigenvalues $\{\lambda_\ell\}_\ell$ of 
the operator $\Theta$ defined in(\ref{DEFo}) which allows us to rewrite (\ref{QUESTAqui0}) as 
\begin{eqnarray} \label{QUESTAqui00}
P_{err,min}^{(n)}  =  \frac{1}{2}\left(1 - {{\sum_\ell}^+}  \lambda_{\ell} \right) \;, \end{eqnarray} 
the sum being restricted  on the positive part of the spectrum.
Since  $\Theta=\alpha^{(n)}-\beta^{(n)}$, we first focus on diagonalizing $\alpha^{(n)}$ and $\beta^{(n)}$ exploiting their symmetry properties. Then, by noticing the common symmetries of $\alpha^{(n)}$ and $\beta^{(n)}$, one can reduce the problem to a diagonalization of $2\times 2$ matrices, as it was already shown in \cite{Sentisprogram}. Here we outline the procedure, which is common to all the scenarios that we consider.
First of all we choose a convenient decomposition of the Hilbert space $\mathcal H_{AXB}$ of the system AXB. 

By Schur-Weyl duality \cite{FultonHarris} the Hilbert space of $n$ multiple qubits can be decomposed as
\be\label{ShWdual2}
\mathcal H= \bigoplus_{D} (j_{D} \otimes \mu_{D})
\ee
  where $j_{D}$ and $\mu_{D}$ are the irreducible representations with Young diagram $D$ respectively of $SU(2)$ and of the symmetric group $S_{n}$.
  
In particular the Hilbert space of the $n$-qubit systems $A$ and $B$ can be expressed as 
\be \label{QUESTAQUI1} 
{\cal H}_A=\oplus_s {\cal H}_{A}^{(s)} \;, \quad {\cal H}_B=\oplus_t {\cal H}_{B}^{(t)} \;, 
\ee
where the labels $s$, $t$ are half-integers varying from $0$ (if $n$ is even) or from $1/2$ (if $n$ is odd) to $n/2$;   
${\cal H}_{A}^{(s)}=\oplus_i {\cal H}_{A_i}^{(s)}$ and ${\cal H}_{B}^{(t)}=\oplus_k {\cal H}_{B_k}^{(t)}$, respectively, are the direct sums of copies of the irreducible representations of $SU(2)$ of dimension $2s+1$ and $2t+1$, while the indexes 
$i$ and $k$ resolve their associated multiplicities, and correspond to a basis of the  irreducible representations of $S_{n}$ associated respectively to $s, t$  -- the multiplicity is given by Eq. (\ref{catalan}) below. 

Accordingly we can then express the joint Hilbert space $\mathcal H_{AXB}={\cal H}_{A}\otimes {\cal H}_X\otimes {\cal H}_B$ of our  $2n+1$ qubits system $AXB$ as the direct sum over $s$, $t$, $i$ and $k$, of the spaces
\be \label{QUESTAQUI1} 
{\cal H}_{A_iXB_k}^{(s,t)}  = {\cal H}_{A_i}^{(s)} \otimes {\cal H}_X  \otimes   {\cal H}_{B_k}^{(t)} \;.
\ee

$\mathcal H_{AXB}$ also carry a representation of $SU(2)$ which sends $U$ to $\hat U^{\otimes 2n+1}$. In particular this representation is reducible and block diagonal in the sectors ${\cal H}_{A_iXB_k}^{(s,t)}$.

In the cases considered, from the symmetry properties of $\alpha^{(n)}$ and $\beta^{(n)}$ one can infer the following symmetries for $\Theta$:
\begin{itemize}
\item $[\Theta, \hat P^{(A)}_\sigma]=[\Theta, \hat P^{(B)}_{\sigma'}]=0$ for every $\hat P^{(A)}_\sigma,\hat P^{(B)}_{\sigma'}$ qubit permutations acting respectively on $\mathcal H _A$ and $\mathcal H _B$.
\item $[\Theta, \hat U^{\otimes 2n+1}]$ for every $U\in SU(2)$.
\end{itemize}

From the first property, $\Theta$ cannot have nonzero matrix elements between states in inequivalent representations of the permutations acting independently on $\mathcal H _A$ and $\mathcal H _B$; besides, the first property also implies that, by Schur's lemma and Schur-Weyl duality (\ref{ShWdual2}) applied to $\cal H_A$ and $\cal H_B$, $\Theta$ is block-diagonal when decomposing ${\cal H}_{A}\otimes {\cal H}_X\otimes {\cal H}_B$ in terms of
the subspaces ${\cal H}_{A_iXB_k}^{(s,t)}$, and its matrix elements do not depend on $i$ and $k$. 

From the second property $\Theta$ is a scalar operator under the action of $SU(2)$, and from Wigner-Eckart theorem \cite{Varshalovich}, the expectation values of $\Theta$ are further constrained to be of the form
\begin{equation}
\bra{l,q,m}\Theta\ket{l',q',m'}=\delta_{q,q'}\delta_{m,m'}\Theta_{l.l'},
\end{equation}
where $\ket{l,q,m}$ are any basis of eigenvectors of the $AXB$-total angular momentum operators ${{\vec J}_{tot}}^2,J_{tot}^{z}$ associated with the full collections of our $2n+1$ spins, i.e.  ${{\vec J}_{tot}}^2\ket{l,q,m}=q(q+1)\ket{l,q,m}$, $J_{tot}^{z}\ket{l,q,m}=m\ket{l,q,m}$. In particular, in each of the ${\cal H}_{A_iXB_k}^{(s,t)}$ blocks the label $q$ span from $||t-s|-1/2|$ to  $t+s+1/2$, while $m$ runs from $q$ to $-q$. $\Theta_{l.l'}$ is  usually called reduced matrix element and depends only on the additional labels $l,l'$.
Putting all together, it follows that for each assigned value of $s$, $t$, $i$, and $k$, 
$\Theta$ further decomposes in a collection of $2\times 2$ or $1\times1$ block diagonal matrices whose elements exhibit functional dependence only on the indexes $s,t,q$.
In particular, by first merging $A_i$ and $X$ and then coupling the two with $B_k$, a convenient orthonormal basis of ${\cal H}_{A_iXB_k}^{(s,t)}$ is provided by the 
following list of vectors 
\bea\{ \ket{s'=s\pm1/2,t; q,m}_{i,k}\}_{q,m} \label{BASE1}  \eea 
   We stress that in the above construction, and in the remaining of the paper,  
   it is implicit assumed that  $\ket{s\pm1/2,t; q,m}_{i,k}$ is null  whenever the parameters $s$, $t$ and $q$ do not fit the necessary angular momentum selection rules.
   This  allows us to identify four different scenarios: 
   \begin{itemize}
   \item[{\it a)}]   $q=s+t+\frac{1}{2}$; 
    \item[{\it b)}]   $q=t-s-\frac{1}{2}$ and  $t>s$; 
    \item[{\it c)}]   $q=s-t-\tfrac{1}{2}$ and  $s>t$;
    \item[{\it d)}] all $s$, $t$, $q$ fitting the selection rules which are not included in the previous cases. 
    \end{itemize} 
 In the first three cases, only one of the elements of the couple $\{ \ket{s\pm1/2,t; q,m}_{i,k}\}$ survives: specifically 
  the  $s+1/2$ element for {\it a)} and {\it b)}, while the $s-1/2$ element for {\it c)}. 
Under such circumstances the symmetry of  $\Theta$ forces it to be  
   $1\times1$ block diagonal, i.e. to 
    admit the associated basis elements   as explicit eigenvectors with eigenvalues $\lambda_{s,t,q}^{(n)}$ that we can formally compute as 
   \begin{equation} 
\Theta_{++}^{(s,t,q)} = _{i,k}\!\bra{s+1/2,t;q,m}\Theta\ket{s+1/2,t;q,m}_{i,k} \;, \label{T++}
 \end{equation}  
   for the case cases  {\it a)} and {\it b)}, and 
   \begin{equation}
  \Theta_{--}^{(s,t,q)}= _{i,k}\!\bra{s-1/2,t;q,m}\Theta\ket{s-1/2,t;q,m}_{i,k} \;,  \label{T--} 
  \end{equation}   
for the {\it c)} case. 
The corresponding  multiplicity is determined instead  by the allowed ranges of $m$, $i$ and $k$, i.e. 
\bea \label{defM}
M_{s,t,q}^{(n)}=(2 q+1) \;  \#(s,n)\; \#(t,n)  \;,
\eea 
with $(2q+1)$ enumerating the possible values of $m$, and with $\#(j,n)$ representing instead the multiplicity of the representations of $SU(2)$ with dimension $2j+1$ in the decomposition of  $n$ spins $1/2$, i.e. 
\bea \label{catalan} 
\#(j,n)&=&
\frac{n!\; (2j+1)}{\left(\frac{n-2j}{2}\right)!\left(\frac{n+2j}{2}+1\right)!}\;. 
\eea

In the scenario  {\it d)} instead 
both the elements of the couple $\{ \ket{s\pm1/2,t; q,m}_{i,k}\}$ survive and the symmetry of the problem forces 
$\Theta$ to be described by 
     $2\times 2$ block diagonal terms ${\Theta}|_{i,k}^{s,t,q,m}$ of the form,
     \bea {\Theta}|_{i,k}^{s,t,q,m} \equiv  \left[  \begin{smallmatrix}
\Theta_{++}^{(s,t,q)} & &\Theta_{+-}^{(s,t,q)}  \\ \\
\Theta_{-+}^{(s,t,q)} & & \Theta_{--}^{(s,t,q)}
\end{smallmatrix}\right]\;, \label{MATRIX} \eea
with $\Theta_{++}^{(s,t,q)}$ and $\Theta_{--}^{(s,t,q)}$ as in (\ref{T++}) and (\ref{T--}) and with 
 \bea 
&\Theta_{+-}^{(s,t,q)} = [\Theta_{-+}^{(s,t,q)}]^* =& \\
&_{i,k}\bra{s+1/2,t;q,m}\Theta\ket{s-1/2,t;q,m}_{i,k} &\;. \nonumber
\label{DEFTHETA} 
\eea
Accordingly we get 
 a further set of  eigenvalues  identified with the functions
 \begin{eqnarray} \label{AUTO}
&&\lambda_{s,t,q}^{(n)}(\pm)    =  \left(\tfrac{\Theta_{--}^{(s,t,q)} +  \Theta_{++}^{(s,t,q)}}{2}\right)  \\
&&  \pm \sqrt{ \left(\tfrac{\Theta_{--}^{(s,t,q)} -  \Theta_{++}^{(s,t,q)}}{2}\right)^2 +  |\Theta_{+-}^{(s,t,q)} |^2}\;, \nn 
\end{eqnarray}
 again characterized by multiplicities  $M^{(n)}_{s,t,q}$ defined as in Eq.~(\ref{defM}). 
 The corresponding eigenvectors are instead  provided by the superpositions 
 \begin{eqnarray} \label{eigvector}
|\psi_{s,t;q,m}^{(\pm)}\rangle_{i,k}&=& 
A^{(s,t,q)}(\pm) \ket{s+1/2,t;q,m}_{i,k}\nn\\
&&+B_{s,t,q}^{(n)}(\pm) \ket{s-1/2,t;q,m}_{i,k}\;, \nn \\
  \end{eqnarray} 
  with amplitudes $A^{(s,t,q)}=\Theta_{+-}^{(s,t,q)}$ and 
 \begin{eqnarray} \label{AUTO}
&&B_{s,t,q}(\pm)    =  \left(\tfrac{\Theta_{--}^{(s,t,q)} -  \Theta_{++}^{(s,t,q)}}{2}\right)  \\
&&  \pm \sqrt{ \left(\tfrac{\Theta_{--}^{(s,t,q)} -  \Theta_{++}^{(s,t,q)}}{2}\right)^2 +  |\Theta_{+-}^{(s,t,q)} |^2}\;, \nn 
\end{eqnarray}
which, for easy of notation we present in a non-normalized form.

\subsection{Scenario i): Mixed states with fixed purity} \label{SEC:i} 
Adopting the Bloch sphere  representation we express the template states  $\rho_1$ and $\rho_2$  in terms of  their associated Bloch vectors $\mathbf r_1$ and $\mathbf r_2$ via the mapping 
\be
\rho_1=\frac{\mathbf 1 +\mathbf r_1 \cdot \sigma}{2},\qquad \rho_2=\frac{\mathbf 1 +\mathbf r_2 \cdot \sigma}{2}\;,
\ee
with $\sigma = (\sigma_x, \sigma_y, \sigma_z)$ being the Pauli vector. 
Assuming then the purity of these density matrices to be assigned, we keep the modulus  $r_1\equiv |\mathbf r_1|$  and $r_2\equiv |\mathbf r_2|$ constant
and use $d\mu(\rho_1,\rho_2)$ to average over all possible  orientations of $\mathbf r_1$ and $\mathbf r_2$ by setting it equal to
\bea 
d\mu(\rho_1,\rho_2) =  dU_1  dU_2\;, \label{MES1} 
\eea  with $dU$ representing  the Haar measure on the unitary transformations of $SU(2)$.
Accordingly we rewrite Eq.~(\ref{DEFalpha})  as  
\bea{\alpha^{(n)}}&=&\int dU_1 \left(U_1\rho_1 U_1^\dagger\right)^{\otimes n+1}\nn\\&&\otimes \int dU_2 \left(U_2\rho_2 U_2^\dagger\right)^{\otimes n},\eea 
\bea{\beta^{(n)}}&=&\int dU_1 \left(U_1\rho_1 U_1^\dagger\right)^{\otimes n}\nn\\&&\otimes\int dU_2 \left(U_2\rho_2 U_2^\dagger\right)^{\otimes n+1}.\label{DEFalpha2} \eea
With this choice both ${\alpha^{(n)}}$ and ${\beta^{(n)}}$, as well as their difference $\Theta$, become explicitly invariant under unitaries acting in the same way on each qubit, i.e. $U^{\otimes 2n +1}$. Therefore the  eigenvectors of each one of these operators must be also eigenvectors of the total angular momentum of the total system $AXB$. 
Furthermore, we notice that 
 ${\alpha^{(n)}}$ and ${\beta^{(n)}}$ are also invariant under separate rotations of partitions of the system, in particular $AX/B$ for ${\alpha^{(n)}}$ and $A/XB$ for ${\beta^{(n)}}$. 
 Following Appendix \ref{APPA1}, for $\rho$ with Bloch vector of modulus $r$ one has the identity
\be \int dU \left(U\rho U^\dagger\right)^{\otimes n}=\oplus_{j}f_j^{(n)}(r)\mathbf 1^{(j)},
\ee
where
 \bea 
f_j^{(n)}(r)&=&\frac{1}{2j+1}\left(\frac{1-r^2}{4}\right)^{\frac n 2-j} \nonumber \\
&\times& \frac{\left(\frac{1+r}{2}\right)^{2j+1}-\left(\frac{1-r}{2}\right)^{2j+1}}{r}\;,  \label{DEFFUNZIONE} 
\eea

This allows us to cast the first of equations (\ref{DEFalpha2}) in the following form 
\bea{\alpha^{(n)}}=\oplus_{s',t}f_{s'}^{(n+1)}(r_1)f_{t}^{(n)}(r_2)\mathbf 1^{(s')}_{AX}\otimes\mathbf 1^{(t)}_{B}\;, 
 \label{operatoraverage}
\eea
where $\mathbf 1^{(j)}_{Q}$ indicates the projector on all the irreducible representations in the system $Q$ with dimension $2j+1$ (i.e. 
 the space ${\cal H}_{B}^{(t)}$ of~(\ref{QUESTAQUI1})  for $\mathbf 1^{(t)}_{B}$, and the irreducible representations of dimension $2s'+1$ in
${\cal H}_{A}^{(s)}\otimes {\cal H}_X$, with $s'=s\pm1/2$ for $\mathbf 1^{(s')}_{AX}$).
Adopting the basis $\{ \ket{s'=s\pm1/2,t; q,m}_{i,k}\}_{q,m}$, defined in ~(\ref{BASE1}) we can then use  Eq.~(\ref{operatoraverage}) to decompose ${\alpha^{(n)}}$ as a direct sum of independent contributions acting on the subspaces ${\cal H}_{A_iXB_k}^{(s,t)}$, i.e. 
\begin{eqnarray}  \label{DEFA} 
{\alpha^{(n)}} = \oplus_{s,t} \oplus_{i,k} (\oplus_{q,m} {\alpha^{(n)}}|_{i,k}^{s,t,q,m}) \;, \eea
where, for each $s,t,i$ and $k$  we exploited the fact that each term further decompose into  a direct sum of either $1\times 1$ or $2\times2$ blocks of the form 
\bea\label{explA}
&&{\alpha^{(n)}}|_{i,k}^{s,t,q,m}=f_{s+1/2}^{(n+1)}(r_1)f_{t}^{(n)}(r_2)  \\&&\ket{s+1/2,t;q,m}_{i,k}\bra{s+1/2,t;q,m}\nn\\&&+f_{s-1/2}^{(n+1)}(r_1)f_{t}^{(n)}(r_2)  \nn\\&&\ket{s-1/2,t;q,m}_{i,k}\bra{s-1/2,t;q,m} \;,\nn 
\eea
where as already mentioned it is implicit assumed that the vectors $\ket{s\pm1/2,t;q,m}_{i,k}$ nullify  whenever the parameters $s,t$ and $q$ do not fit the angular momentum selection rules.
In a similar fashion we have that 
\bea
\label{DEFB} 
{\beta^{(n)}}=\oplus_{s,t'}f_{s}^{(n)}(r_1)f_{t'}^{(n+1)}(r_2)\mathbf 1^{(s)}_{A}\otimes\mathbf 1^{(t')}_{XB}\;, \eea 
where now $\mathbf 1^{(s)}_{A}$ project on 
 ${\cal H}_{A}^{(s)}$ of~(\ref{QUESTAQUI1})  and $\mathbf 1^{(t')}_{XB}$  on the irreducible representations $t'$ in 
${\cal H}_{X}\otimes {\cal H}_B^{(t)}$, $t'=t\pm1/2$. Again this yields the following decomposition 
\begin{eqnarray} \label{DEFB} 
{\beta^{(n)}} =   \oplus_{s,t} \oplus_{i,k} \left( \oplus_{q,m}  {{\beta^{(n)}}}|_{i,k}^{s,t,q,m}\right) \;, \eea
where now
\bea\label{explB}
&&{{\beta^{(n)}}}|_{i,k}^{s,t,q,m}=f_{s}^{(n)}(r_1)f_{t+1/2}^{(n+1)}(r_2)\\&&\ket{s,t+1/2;q,m}_{i,k}\bra{s,t+1/2;q,m}\nn\\&&+f_{s}^{(n)}(r_1)f_{t-1/2}^{(n+1)}(r_2)\nn\\&&\ket{s,t-1/2;q,m}_{i,k}\bra{s,t-1/2;q,m}\;. \nn 
\eea
In this expression the elements 
\begin{eqnarray} \{ \ket{s,t'=t\pm1/2; q,m}_{i,k}\}_{q,m}\label{BASE2} \;, 
\end{eqnarray}   are obtained by 
coupling the qubit  ${\cal H}_X$ with those of ${\cal H}^{(t)}_{B,k}$ and, as usual, we assume they nullify whenever  $s,t$ and $q$ do not
fulfil the necessary selection rules. These vectors form a new basis for ${\cal H}_{A_iXB_k}^{(s,t)}$  connected with $\{ \ket{s'=s\pm1/2,t; q,m}_{i,k}\}_{q,m}$ via the 
following four amplitude probabilities 
\bea
C^{(s,t,q)}_{++}\equiv{_{i,k}}\braket{s+\tfrac12,t;q,m}{s,t+\tfrac 1 2;q,m}_{i,k}\;, \nn\;&&\\
C^{(s,t,q)}_{+-}\equiv{_{i,k}}\braket{s+\tfrac 1 2,t;q,m}{s,t-\tfrac 1 2;q,m}_{i,k} \;,\nn\;&&\\
C^{(s,t,q)}_{-+}\equiv{_{i,k}}\braket{s-\tfrac 1 2,t;q,m}{s,t+\tfrac 1 2;q,m}_{i,k}\;,\nn\;&&\\
C^{(s,t,q)}_{--}\equiv{_{i,k}}\braket{s-\tfrac 1 2,t;q,m}{s,t-\tfrac 1 2;q,m}_{i,k} \;, &&\label{CCC}
\eea
which express a unitary transformation between the two different recouplings (\ref{BASE1}) and (\ref{BASE2}) of the irreducible representations $s,t,\frac1 2$. This is exactly the information that the Wigner 6j symbols \cite{Varshalovich} of $SU(2)$ encode, and indeed $C^{(s,t,q)}_{\pm\pm}$ can be written as
\bea 
C^{(s,t,q)}_{\pm\pm}&=&(-1)^{\pm \frac 1 2 \pm\frac 1 2}\nn\\
&\times&\sqrt{(2s\pm 1+1)(2t\pm 1+1)}\nn\\&\times& \left\{
\begin{matrix}
t\pm \frac 1 2 &t& \frac 1 2\\
s\pm \frac 1 2 &s& q
\end{matrix} \label{WIG} 
\right\}\;,
\eea
which for the particular case at hand gives a closed analytic expression. 
Notice that $C^{(s,t,q)}_{\pm\pm}$ do not depend on $m$, by virtue of Wigner-Eckart theorem, since the unitary transformation that they define commutes with the action of $SU(2)$ on the space ${\cal H}_{A_iXB_k}^{(s,t)}$. 
 
From Eqs.~(\ref{DEFA}) and (\ref{DEFB}) it now follows that a similar decomposition holds also for $\Theta$,
\begin{eqnarray} 
\Theta = \oplus_{s,t} \oplus_{i,k}\left( \oplus_{q,m} {\Theta}|_{i,k}^{s,t,q,m} \right), \eea
where for assigned $s,t,i$ and $k$, ${\Theta}|_{i,k}^{s,t,q,m}$ are the  following $1\times 1$ or  $2\times2$ matrices  
 \bea\label{DDD} 
{\Theta}|_{i,k}^{s,t,q,m} = {{\alpha^{(n)}}}|_{i,k}^{s,t,q,m} - 
{{\beta^{(n)}}}|_{i,k}^{s,t,q,m}\;.
\eea
Invoking the convention  established when introducing  Eq.~(\ref{BASE1}) we notice that $1\times 1$ blocks 
occur explicitly in the scenarios detailed in the introductory part of  the section: {\it a)} $q=s+t+\frac{1}{2}$, {\it b)} $q=t-s-\frac{1}{2}$ and  $t>s$, and {\it c)} $q=s-t-\tfrac{1}{2}$ and  $s>t$, yielding 
the  eigenvalues 
\bea \label{RESCALED} 
\lambda_{s,t,q}^{(n)}&=&f_{s}^{(n)}(r_1)f_{t}^{(n)}(r_2)  \; \Lambda_{s,t,q}^{(n)}\;,
\eea
with 
\bea\label{eigmix1}  \Lambda_{s,t,q}^{(n)}  =  \left\{ \begin{array}{ll} 
R_{s,+}^{(n)}(r_1)-R_{t,+}^{(n)}(r_2) & \mbox{case {\it a)},} \\ \\
R_{s,+}^{(n)}(r_1)-R_{t,-}^{(n)}(r_2)  & \mbox{case {\it b)},} \\ \\
R_{s,-}^{(n)}(r_1)-R_{t,+}^{(n)}(r_2)  & \mbox{case {\it c)},} \\
 \end{array} 
\right.   \label{DEFere}   \eea
where we introduced the functions 
\be
R^{(n)}_{j,\pm}(r)\equiv \frac{f_{j\pm1/2}^{(n+1)}(r)}{f_{j}^{(n)}(r)}\;.
\ee
For $s$, $t$, and $q$ belonging to the remaining case {\it d)} instead, (\ref{DDD}) is a $2\times2$ matrix of the form 
 (\ref{MATRIX})
$$ \left[  \begin{smallmatrix}
\Theta_{++}^{(s,t,q)} & &\Theta_{+-}^{(s,t,q)}  \\ \\
\Theta_{-+}^{(s,t,q)} & & \Theta_{--}^{(s,t,q)}
\end{smallmatrix}\right],$$
with eigenvalues as in  (\ref{AUTO}) with the following identifications 
\begin{eqnarray} 
&&\Theta^{(s,t,q)}_{++}=f_{s}^{(n)}(r_1)f_{t}^{(n)}(r_2)  \left[ R_{s,+}^{(n)}(r_1)\right.  \nonumber \\ && \quad \left. -  R_{t,+}^{(n)}(r_2)(C^{(s,t,q)}_{++})^2-R_{t,-}^{(n)}(r_2)(C^{(s,t,q)}_{+-})^2 \right],\nn
\\
&&\Theta^{(s,t,q)}_{--}=f_{s}^{(n)}(r_1)f_{t}^{(n)}(r_2)  \left[ R_{s,-}^{(n)}(r_1)\right.  \nonumber \\ &&  \quad  \left. -R_{t,+}^{(n)}(r_2)(C^{(s,t,q)}_{-+})^2-R_{t,-}^{(n)}(r_2)(C^{(s,t,q)}_{--})^2  \right],  \nn
\eea
and 
\begin{eqnarray}
&&\Theta^{(s,t,q)}_{+-}=- f_{s}^{(n)}(r_1)f_{t}^{(n)}(r_2)\nn \\ 
&& \!\!\!\!\!\!\!\!\! \!\!\!\!\!\! \left[ R_{t,+}^{(n)}(r_2)C^{(s,t,q)}_{++}C^{(s,t,q)}_{-+}+
  R_{t,-}^{(n)}(r_2)C^{(s,t,q)}_{+-}C^{(s,t,q)}_{--} \right] ,  \nn
\end{eqnarray} 
where we used the  coefficients $C^{(s,t,q)}_{\pm\pm}$ (\ref{CCC}) to express the elements of ${{\beta^{(n)}}}|_{i,k}^{s,t,q,m}$ into the basis $\{ \ket{s'=s\pm1/2,t; q,m}_{i,k}\}_{q,m}$.
The corresponding eigenvalues can also be expressed as in the rescaled form (\ref{RESCALED}) with 
\begin{eqnarray}\label{RESCALEDEI|}
\Lambda_{s,t,q}^{(n)} (\pm) =  a_{s,t}^{(n)} \pm b_{s,t,q}^{(n)}\;,
 \end{eqnarray} 
the functions   $a^{s,t}$ and $b^{s,t}_q$ being defined as 
\be
a_{s,t}^{(n)}\equiv  \tfrac{R_{s,+}^{(n)}(r_1)+ R_{s,-}^{(n)}(r_1) -R_{t,+}^{(n)}(r_2)-R_{t,-}^{(n)}(r_2)}{2}\;, \ee
\be b_{s,t,q}^{(n)} \equiv \tfrac{\sqrt{[G_{s}(r_1)-G_{t}(r_2)]^2  -4G_{s}(r_1)G_{t}(r_2){(C^{(s,t,q)}_{++})^2}}}{2}\;, 
\ee
where for ease of notation  we introduced 
\bea
G_{j}(r)&\equiv& f_{j+1/2}^{(n+1)}(r)- f_{j-1/2}^{(n+1)}(r)\;.
\eea
For future reference we  observe that from Eq.~(\ref{WIG}) the following inequality can be determined 
\bea
b_{s,t,q}^{(n)} \geq  b_{s,t,q=s+t-1/2}^{(n)}\;, 
\eea
which in turn can be used to establish  useful  bounds for  the eigenvalues (\ref{RESCALEDEI|}), i.e. 
\begin{eqnarray} 
\Lambda_{s,t,q}^{(n)} (+) \geq  \Lambda_{s,t,q=s+t-1/2}^{(n)} (+)  \;,  \label{DD1} \\
\Lambda_{s,t,q}^{(n)}(-) \leq  \Lambda_{s,t,q=s+t-1/2}^{(n)} (-)  \;.\label{DD2}
 \end{eqnarray}

 Replacing all this into Eq.~(\ref{QUESTAqui00}) we can finally write 
 \begin{eqnarray} \label{QUESTAqui0000011}
 &&P_{err,min}^{(n)}  = \frac{1}{2}  \\\nn
&& \,- \frac{1}2 {\sum_{s,t,q,\ell }}^+ f_{s}^{(n)}(r_1)f_{t}^{(n)}(r_2)\;  M_{s,t,q}^{(n)}\; {\Lambda}_{s,t,q}^{(n)}(\ell) 
 \;, \end{eqnarray} 
with $M_{s,t,q}^{(n)}$ being the multiplicity factor defined in Eq.~(\ref{defM}),  the index $\ell$ assuming the values $\pm$ for the case {\it d)}, and  
 where the subscript $^+$ indicates that only the  positive values of ${\Lambda}_{s,t,q}^{(n)}(\ell)$ are allowed into the sum. 
In order to get an asymptotic expansion of Eq.~(\ref{QUESTAqui0000011})  we now notice that for large $n$ the following expansion holds, 
\bea\label{approx}
f_{s}^{(n)}(r)\#(s,n)\approx \frac {1+r} {r} \frac {1}{1+\frac n 2 +s}\nn\\ \times B(n,\frac{1+r}{2},n/2+s)
\eea
where $B(n,\frac{1+r}{2},n/2+s)$ is a binomial distribution for the variable $n/2+s$, and the neglected terms give an exponentially suppressed contribution as $n$ goes to infinity. The mean of $\frac s n $ is $ \frac{r}{2}$ and the variance is $ \frac{1-r^2}{4 n}$, the next moments give contribution $O(n^{-2})$. The sum on $s$ goes from zero or $1/2$ to $n/2$, therefore if $r$ is sufficiently greater than $0$ we are neglecting in the sum a region where the binomial distribution in small and the total contribution of the region to the sum is exponentially suppressed. The second useful observation is that the eigenvalues and the term outside the binomial in (\ref{approx}),  expanded in the variables $\frac s n$ and $\frac t n $ around their means, show series coefficients that do not increase in powers of $n$ as one goes to higher terms. Therefore to get the leading and next to leading term one needs the expansion only at second order in these variables.

The expansion in $\frac s n,\frac t n$ around their means let us also determine the sign of the eigenvalues in the relevant region for the sum.
In particular for the four cases analyzed so far we have: 

\begin{flushleft}
\begin{itemize}\label{sign}
\item[{\it a)}] $\Lambda_{s,t,q=s+t+1/2}^{(n)}=\frac{r_1-r_2}{2}+O\left(|\frac s n-\frac{r_1}{2}|+|\frac t n-\frac {r_2}{2}|+|\frac 1 n|\right)$,
\item[{\it b)}] $\Lambda_{s,t,q=t-s-1/2}^{(n)}=\frac{r_1+r_2}{2}+O\left(|\frac s n-\frac{r_1}{2}|+|\frac t n-\frac {r_2}{2}|+|\frac 1 n|\right)$, 
\item[{\it c)}]  $\Lambda_{s,t,q=s-t-1/2}^{(n)}=-\frac{r_1+r_2}{2}+O\left(|\frac s n-\frac{r_1}{2}|+|\frac t n-\frac {r_2}{2}|+|\frac 1 n|\right)$, 
\item[{\it d)}] 
 $\Lambda_{s,t,q}^{(n)} (+) 
\geq
 \tfrac{\sqrt{(r_1 - r_2)^2}}{2}+O\left(|\frac s n-\frac{r_1}{2}|+|\frac t n-\frac {r_2}{2}|+|\frac 1 n|\right)$,
 $\Lambda_{s,t,q}^{(n)} (-) 
\leq
-\tfrac{\sqrt{(r_1 - r_2)^2}}{2}+O\left(|\frac s n-\frac{r_1}{2}|+|\frac t n-\frac {r_2}{2}|+|\frac 1 n|\right)$,

as $\frac{s}{n}\rightarrow \frac {r_1} 2$, $\frac{t}{n}\rightarrow \frac { r_2} 2$, and $n\rightarrow \infty$.
\end{itemize}
\end{flushleft}
where in deriving the last two inequalities we used~(\ref{DD1}) and (\ref{DD2}). 
The above expressions allows us to identify the 
positive terms which, in the limit of large $n$, contribute to the sum (\ref{QUESTAqui0000011}): for instance 
taking $r_1>r_2$ we noticed that the positive eigenvalues are those associated with case {\it a)} and the first of case {\it d)}, while the case {\it b)}, 
 which is also positive,
 can be ignored because $t>s$ is not in the relevant region of the sum on $s, t$. 
With this information, the sum on $q$ can now be performed at the relevant order with the second order of the Euler-MacLaurin expansion (the details are available in the supplementary Mathematica \cite{Mathematica} notebooks, available at \cite{github}):
  \be\label{McL}
\sum_{i=a}^b f(i)\approx \int_a^b f(x) dx +\frac{f(a)+f(b)}{2}.
\ee

The final result, which takes into account also the case $r_1<r_2$, is
\bea\label{perrmix}
P_{err,min}^{(n\gg1)} &\simeq &\frac 1 2 -\frac 1 {24} \frac{(r_1+r_2)^3-|r_1-r_2|^3}{r_1 r_2}\nn\\
&+&\frac 5 {24 \,n}\frac{(r_1+r_2)^3+|r_1-r_2|^{3}}{ r_1^2 r_2^2}\nn\\ 
&-&\frac 1 {24 \,n}\frac{(r_1+r_2)^5-|r_1-r_2|^5}{r_1^3 r_2^3 }.
\eea
which for  $r_1=r_2$  reproduce correctly the result of~\cite{Sentisprogram}, 
and which in agreement with~(\ref{QUESTAqui}) exhibits a leading order that corresponds to the average of the Helstrom probabilities, i.e. 
\bea \label{FFD} 
\bar{P}_H&=&\frac 1 2 -\frac 1 4 \int \sin \theta d\theta \tfrac{\sqrt{(r_1 - r_2 \cos\theta)^2 + r_2\sin^2\theta}}{2}\nn\\&=&\frac 1 2 -\frac 1 {24} \frac{(r_1+r_2)^3-|r_1-r_2|^3}{r_1 r_2}\;.\nn\\
\eea
 In Figure \ref{fixpur} we show the comparison between the exact values of {$P_{err,min}^{(n)}$ (\ref{QUESTAqui0000011})  and the asymptotic expansion~(\ref{perrmix}). }

\begin{figure}
    \centering
        \includegraphics[width=0.47\textwidth]{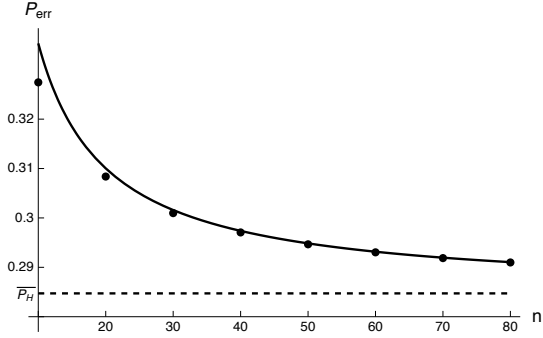}
\caption{\label{fixpur}{\small{\emph{Scenario i) Minimal probability of error as a function of $n$, with $r_1=\frac 3 4$ and $r_2=\frac 1 2$: exact values (dots), asymptotic expansion Eq. (\ref{perrmix}) (solid line), Helstrom probability (dashed line).} }} }
\end{figure}

\subsection{Scenario ii): Mixed states with hard sphere prior} \label{SEC:ii}

In the scenario ii) we are interested in considering the case where $\rho_1$ and $\rho_2$ are arbitrary (possibily) mixed density matrices.
This corresponds to replace (\ref{MES1}) with 
\bea 
d\mu(\rho_1,\rho_2) =  dU_1 d\mu(r_1)  dU_2 d\mu(r_2) \;, \label{MES2} 
\eea 
where again $dU$ represents the Haar measure of $SU(2)$ while $d\mu(r)$ is a measure that gauges our ignorance about the purity of the template states, i.e. the length of their associated Bloch vectors. 
Accordingly the only difference with the previous paragraph is that now, in the expression of  ${\alpha^{(n)}}= \oplus_{s,t} \oplus_{i,k} (\oplus_{q,m} {\alpha^{(n)}}|_{i,k}^{s,t,q,m})$ and ${\beta^{(n)}}=  \oplus_{s,t} \oplus_{i,k} \left( \oplus_{q,m}  {{\beta^{(n)}}}|_{i,k}^{s,t,q,m}\right)$ given in Eqs.~(\ref{explA}) and (\ref{explB})   we have now  to replace the functions   $f_j^{(n)}(r)$ with their averaged values, i.e. 
\be
f_j^{(n)}(r) \rightarrow f_j^{(n)} \equiv \int  d\mu(r) f_j^{(n)}(r) \;, 
\ee
such that
\bea\label{explA}
&&{\alpha^{(n)}}|_{i,k}^{s,t,q,m}=f_{s+1/2}^{(n+1)}f_{t}^{(n)}  \\&&\ket{s+1/2,t;q,m}_{i,k}\bra{s+1/2,t;q,m}\nn\\&&+f_{s-1/2}^{(n+1)}f_{t}^{(n)} \nn\\&&\ket{s-1/2,t;q,m}_{i,k}\bra{s-1/2,t;q,m} \;,\nn 
\eea
and
\bea\label{explB}
&&{{\beta^{(n)}}}|_{i,k}^{s,t,q,m}=f_{s}^{(n)}f_{t+1/2}^{(n+1)}\\&&\ket{s,t+1/2;q,m}_{i,k}\bra{s,t+1/2;q,m}\nn\\&&+f_{s}^{(n)}f_{t-1/2}^{(n+1)}\nn\\&&\ket{s,t-1/2;q,m}_{i,k}\bra{s,t-1/2;q,m}\;. \nn 
\eea
As a choice for $d\mu(r)$ we take the hard sphere prior measure, i.e. 
\be d\mu(r) = 3 r^2 dr\ee
which yields 
\be
f_j^{(n)} =6\frac{\left(\frac n 2 -j\right)!\left(1+\frac n 2 +j\right)!}{(n+3)!}\;.
\ee
The associated eigenvalues of $\Theta$ can then be expressed as in (\ref{RESCALED}) with the rescaled quantities $\Lambda_{s,t,q}^{(n)}$ such that the eigenvalues $\lambda_{s,t,q}^{(n)}$ are
$$\lambda_{s,t,q}^{(n)}=f_{s}^{(n)}f_{t}^{(n)} \; \Lambda_{s,t,q}^{(n)}\;,$$
$\Lambda_{s,t,q}^{(n)}$ are obtained as in (\ref{eigmix1}),(\ref{RESCALEDEI|}), with the terms
$R^{(n)}_{s,\pm}(r)$ being replaced by 
\bea
&&R^{(n)}_{s,+}\equiv \frac{f_{s+1/2}^{(n+1)}}{f_{s}^{(n)}}=\frac{2+\frac{n}{2}+s}{n+4}, \nn\\&& R^{(n)}_{s,-}\equiv \frac{f_{s-1/2}^{(n+1)}}{f_{s}^{(n)}}=\frac{1+\frac{n}{2}-s}{n+4}\;,
\eea
and the same for $R^{(n)}_{t,\pm}(r)$ .

As a result,  for the cases {\it a)}, {\it b)}, {\it c)}, and {\it d)}, we get the following solutions, 
\begin{flushleft}
\begin{itemize}\label{sign}
\item[{\it a)}] $\Lambda^{(n)}_{s,t,q = s+t+\tfrac 1 2}=\frac{s-t}{n+4}$,
\item[{\it b)}] $\Lambda^{(n)}_{s,t,q = t-s-\tfrac 1 2}=\frac{1+s+t}{n+4}$,
\item[{\it c)}]  $\Lambda^{(n)}_{s,t,q = s-t-\tfrac 1 2}=-\frac{1+s+t}{n+4}$. 
\item[{\it d)}] 
 $\Lambda^{(n)}_{s,t,q}(\pm) 
=\pm \frac {\sqrt{3 - 4 q (1 + q) + 8 s (1 + s) + 8 t (1 + t)}}{2(n+4)}$,
\end{itemize}
\end{flushleft}
which shows that only terms entering   in the expression (\ref{QUESTAqui0000011}) for $P_{err,min}^{(n)}$ are those of  {\it a)} with $s>t$, those of {\it b)}, and the $\Lambda^{(n)}_{s,t,q}(+)$ term of   {\it d)}.
Accordingly we can write  
 \begin{eqnarray} \label{QUESTAqui000001331}
 && P_{err,min}^{(n)}  = \frac{1-S^{(n)}}{2}  
 \;, \end{eqnarray} 
with 
\bea
S^{(n)}=\sum_{s>t} f_s^{(n)} f_t^{(n)}M_{s,t,s+t+\tfrac 1 2}^{(n)}\; \Lambda^{(n)}_{s,t,s+t+\tfrac 1 2}
  &\nn\\
+ \sum_{t>s}f_s^{(n)} f_t^{(n)}  M_{s,t,t-s-\tfrac 1 2}^{(n)}  \;   \Lambda^{(n)}_{s,t,t-s-\tfrac 1 2}&\nn\\
+\sum_{s,t}f_s^{(n)} f_t^{(n)}\sum_{q=|s-t|+\tfrac 1 2}^{s+t-\tfrac 1 2} M_{s,t,q}^{(n)} \;\Lambda^{(n)}_{s,t,q}(+),&\nn\\
\eea
with $M_{s,t,q}^{(n)}$ the multiplicity factors of defined in Eq.~(\ref{defM}) which allow
for a simplification of the resulting formula thanks to the   identity
\be
f_s^{(n)}\;  f_t^{(n)}  M_{s,t,q}^{(n)}= \tfrac{36(2s+1)(2t+1)(2q+1)}{(n+1)^2(n+2)^2}\;.
\ee
To get to the final result at order $O\left(\frac 1 n\right)$ one can still exploit the Euler McLaurin formula (\ref{McL}) for each of the three sums, 
and the details are available in the supplementary Mathematica notebooks.
The result is
\be\label{aveasy}
P_{err,min}^{(n\gg1)} \simeq \frac {17} {70}+\frac{18}{35n}\;,
\ee
which in $n\rightarrow \infty$ agrees with the average Helstrom probability $\bar{P}_H=17/70$ that in the present case can be obtained by integrating (\ref{FFD}) with respect to $r_1$ and $r_2$ with the corresponding hard sphere 
measures. In Figure \ref{aveplot} we show the comparison between the exact values
of $P_{err,min}^{(n)}$ (\ref{QUESTAqui000001331})   and the asymptotic expansion~(\ref{aveasy}).

\begin{figure}
    \centering
        \includegraphics[width=0.47\textwidth]{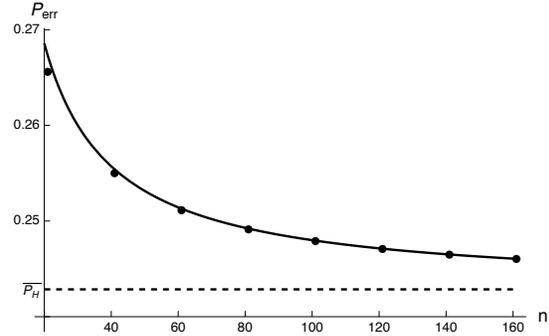}
\caption{\label{aveplot} {\small{\emph{Scenario ii) Minimal probability of error as a function of $n$: exact values (dots), asymptotic expansion Eq. (\ref{aveasy}) (solid line), Helstrom probability (dashed line).} }} }
\end{figure}

\subsection{Scenario iii): Pure states at fixed overlap} \label{SEC:iii} 

We now consider the case where the templates states $\rho_1=|\psi_1\rangle\langle \psi_1|$ and $\rho_2=|\psi_2\rangle\langle \psi_2|$ are pure  and characterized by a mutual overlap which is known a priori. No information about the absolute orientation of the  couple is
instead assumed. As anticipated in the introductory section this model  appears to be  well suited to characterize a scenario where for instance the machine is asked to discriminate between two possible configurations on the basis of templates  
generated by an external party which does not share a common reference frame with the machine itself. 
Without loss of generality we can model this problem by setting 
\begin{eqnarray}
|\psi_1\rangle =\ket{\uparrow}\;, \qquad 
\ket{\psi_2}= U_0\ket{\uparrow}\;, 
\end{eqnarray} 
with a fixed unitary $U_0$, and then average on the action of a unitary transformation $U$ on both $\rho_1,\rho_2$. With this choice the states (\ref{DEFalpha})  become  
\bea\label{alphaoverlap}
{\alpha^{(n)}}=\int dU\left({U\ket{\uparrow}\bra{\uparrow} U^\dagger}\right)^{\otimes n+1} \!\!\nn\\\otimes \left({U U_0\ket{\uparrow}\bra{\uparrow} U_0^\dagger U^{\dagger}}\right)^{\otimes n}
\eea
\bea\label{betaoverlap}{\beta^{(n)}}=\int dU\left({U\ket{\uparrow}\bra{\uparrow} U^\dagger}\right)^{\otimes n} \!\! \nn\\\otimes \left({U U_0\ket{\uparrow}\bra{\uparrow} U_0^\dagger U^{\dagger}}\right)^{\otimes n+1}
\eea
where once more $dU$ is the Haar measure of $SU(2)$. 
As in the cases analyzed before
${\alpha^{(n)}}$ and ${\beta^{(n)}}$, as well as their difference $\Theta^{(n)}$ are invariant under $U^{\otimes 2n+1}$. 
Furthermore,  since on both the $AX$   and the $B$ partition   $\alpha^{(n)}$ is described by pure vectors which are completely symmetric under permutations,  
the only elements of the basis (\ref{BASE1}) on which it can have support  are those with maximum values of $s'$ and $t$, i.e. $s'= (n+1)/2$, $t=n/2$.
As explicitly derived in (\ref{APPAB}), these states are also eigenstates for  $\alpha^{(n)}$, i.e. 
\be
{\alpha^{(n)}}\ket{\tfrac{n+1}2, \tfrac{n}2; q,m}= \Phi^{(n)}(q,U_0)\ket{\tfrac{n+1}2, \tfrac{n}2; q,m}, \label{fixangle}
\ee
with eigenvalues $ \Phi^{(n)}(q,U_0)$ given by  
\bea \nn 
  \Phi^{(n)}(q,U_0)=\sum_{h=-{n}/2}^{{n}/2} D^{\frac{n}{2}}_{h, \frac{n}{2}}(U_0) {D^{\frac{n}{2}}_{\frac{n}{2},h}}(U_0^\dagger) \nn\\\times \frac {1}{2q+1} \; C^{q,\frac {n+1} 2+h}_{\frac{n+1}{2},\frac{n+1}{2},\frac{n}{2},h} C^{q,\frac {n+1} 2+h}_{\frac{n+1}{2},\frac{n+1}{2},\frac{n}{2},h}, \nn  \\ \label{fixangle} \eea
where the symbol  $D^{j}_{m m'}(U)$ represent the matrix elements of the irreducible representations of $U\in SU(2)$ with dimension $2j+1$, and $C^{q,l}_{j ,m, j', m'}$ being the Clebsch-Gordan coefficients.
(Notice that in the above analysis we dropped the multiplicity labels $i$ and $k$ in writing the elements of the basis (\ref{BASE1})  because for the  $s'= (n+1)/2$, $t=n/2$ no degeneracy of the representation is present, $\#(n/2,n) =1$).
Analogous  properties applies for ${\beta^{(n)}}$ when expressed into the basis (\ref{BASE2}).  
Therefore as in the previous cases $\Theta$ can be expressed as a direct sum of $1\times 1$ and $2\times 2$ block matrices. In the present case, however due to the special restriction on $s$ and $t$  instead of the four possible cases observed in the previous section, only {\it a)} and {\it d)} may occur. 
It turns out that for the case {\it a)} the associated eigenvalues is always null. For {\it d)} instead we have 
\bea\label{EIGO} 
\lambda^{(n)}_{s=n/2,t=n/2,q}(\pm) =\pm  \Phi^{(n)}(q,U_0)\qquad\qquad && \\ \nn 
\times |C^{(s=n/2,t=n/2,q)}_{+-}|,&&
\eea
and the eigenvectors are the same that we obtain for $r_1=r_2=1$ in the case of completely random orientations: for pure states, the optimal POVM in the fixed overlap case is the same. Therefore, writing the eigenvalues in a simpler notation as $\lambda_q^{(n)}(\pm)$, we have
\be\label{overlaptheta}
\Theta^{(n)}=\sum_q \left(\lambda^{(n)}_{q}(+)\Pi_{q,+}+\lambda^{(n)}_{q}(-)\Pi_{q,-}\right),
\ee
where $\Pi_{q+}$ and $\Pi_{q-}$ are the projectors on eigenvectors with total angular momentum $q$ and respectively positive and negative eigenvalues.

Replacing all this into Eq.~(\ref{QUESTAqui00}) we can finally write 
 \begin{equation} \label{QUESTAqui0000011sss}
 P_{err,min}^{(n)}  = \frac{1}{2}  - \frac{1}2 {\sum_{q }} (2q+1) \; \lambda^{(n)}_{q}(+)
 \;, \end{equation} 
where we used the fact that $M_{s=n/2,t=n/2,q}^{(n)} = 2q+1$ and that only the $+$ elements of the couples~(\ref{EIGO}) are positive. 
To proceed further, without loss of generality, we write $U_0=\exp(-i \sigma_y (\pi-\theta)/2)$  obtaining
\bea \label{BINOMIAL} 
&D^{\frac{n}{2}}_{h, \frac{n}{2}}(U_0)D^{\frac{n}{2}}_{\frac{n}{2}, h}(U_0^\dagger)=\frac{n!}{\left(\frac {n} 2 +h\right)!\left(\frac {n} 2 -h\right)!}&\\  &\qquad \times \left(\cos^2 \left(\frac{\pi-\theta}{2}\right)\right)^{\frac{n}{2}+h}\left(\sin^2 \left(\frac{\pi-\theta}{2}\right)\right)^{\frac{n}{2}-h}, &\nn 
\eea
which is  a binomial distribution in the variable $\frac n 2 + h \in\{ 0, n\}$.
We also notice  that 
\bea \label{NEWDIST}
&\left(C^{q,\frac {n+1} 2+h}_{\frac{n+1}{2},\frac{n+1}{2},\frac{n}{2},h}\right)^2=\frac{2 \left(\frac n 2 -h\right)!\left(n+1\right)!}{
\left(\frac n 2 +h\right)!}\nn&\\&\times\frac{\left(\frac n 2 +h+q+\frac 1 2\right)!}{
\left(q-\frac 1 2-\frac n 2 -h\right)!\left(n-q+\frac 1 2\right)!\left(n+q+\frac 3 2\right)!},
\eea
is also a probability distribution in the variable $q \in \{ \frac n 2+h, n+\frac 1 2\}$. 
Then the terms entering in the sum of Eq.~(\ref{QUESTAqui0000011sss}) rewrite explicitly as 
\bea
&(2q+1)\lambda^{(n)}_{q}(+)=\sum_h \frac{n!}{\left(\frac {n} 2 +h\right)!\left(\frac {n} 2 -h\right)!}&\nn\\
&\times \left(\cos^2 \left(\frac{\pi-\theta}{2}\right)\right)^{\frac{n}{2}+h}\left(\sin^2 \left(\frac{\pi-\theta}{2}\right)\right)^{\frac{n}{2}-h} \nn&\\& \times 
 \frac{2 \left(\frac n 2 -h\right)!\left(\frac n 2 +h+q+\frac 1 2\right)!\left(n+1\right)!}{
\left(\frac n 2 -h+q-\frac 1 2\right)!\left(\frac n 2 +h\right)!\left(n-q+\frac 1 2\right)!\left(n+q+\frac 3 2\right)!}\nn&\\& 
\times \frac 1 2 \sqrt{\frac{2 (3/2 + q + n) (1/2 - q + n)}{(n/2 + 1/2) (n + 1)}}\;.  
\eea
 As usual we focus  on the limit of large $n\gg1$ for  $P_{err,min}^{(n)}$. In this case we notice that in order to get up to  the order $O(\frac 1 {n^2})$ for the resulting expression, one can expand $|C^{(s=n/2,t=n/2,q)}_{+-}|$ around the mean of the $q$ distribution and consider contributions up to the fourth central moment (see Appendix~\ref{APPAC} and supplementary Mathematica notebooks), expand the result around the mean of the $h$ distribution and calculate the contributions up to the relevant moment (not more than the fourth).
The result is
\bea\label{resultoverlap}
 P_{err,min}^{(n\gg1)}  \simeq\frac 1 2\left(1-|\cos\tfrac \theta 2|\right)+\tfrac{3 + \cos\theta}{8 \sqrt 2 \sqrt{1 + \cos\theta}}\frac{1}{n}\nn\\+\tfrac{1 -60 \cos\theta-5\cos2\theta}{128 \sqrt 2 (1 + \cos\theta)^{3/2}}\frac{1}{n^2} \;, && \label{fixresult}
\eea
where, as expected, the first contribution corresponds to the corresponding averaged Helstrom probability $\bar{P}_H$ -- see also 
Figure ~\ref{FigLOIntro}. 
We notice that for small deviations from orthogonality, one has
\be
P_{err,min}^{(n\gg1)} \simeq \frac{\theta^2}{16} + \frac{1}{4 n} - \frac{1}
 {8 n^2}\left(1-\frac{\theta^2}{4}\right)\;,\ee
The expansion around coincident states is instead singular, but the formula is still valid when the states are not coincident and $n (\pi-\theta)\gg1$. Since the optimal POVM is the same of the totally random pure state scenario, averaging over $\theta$ before doing the asymptotic expansion gives the result of (\ref{perrmix}) when $r_1=r_2=1$. Integrating at the end gives also the same result up to first order, while the order ${n^{-2}}$ is not integrable. This is not inconsistent: one can see that the averaged $P^{(n)}_{err,min}$ displays a ${n^{-\frac 3 2}}$ dependence which is not recoverable from this expansion (and also not exactly computable with the Euler-MacLaurin approximation), which at fixed $n$ works only in the region $n (\pi-\theta)\gg1$. 

\begin{figure}
    \centering
        \includegraphics[width=0.47\textwidth]{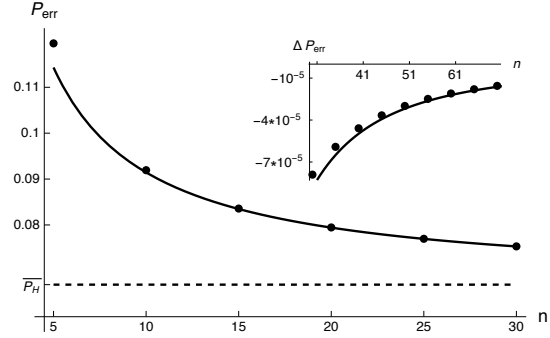}
\caption{\label{FigLOIntro} {\small{\emph{Scenario iii) Minimal probability of error as a function of $n$,with $\theta= \frac \pi 3$: exact values (dots), asymptotic expansion 
Eq.~(\ref{resultoverlap})}} (solid line), Helstrom probability (dashed line). In the inset we show the second order correction.} }
\end{figure}

{\subsubsection{Generalisation in finite dimension $d$} For the special setting of the present scenario we present also a generalization to higher
dimension. For this purpose we now consider $\ket{\psi_1}$ and $\ket{\psi_2}$ as states in a $d$ dimensional Hilbert space. Accordingly, substituting $\ket{\uparrow}$ with $\ket{e}$, we can write the analogue of (\ref{alphaoverlap}), (\ref{betaoverlap}) as
\bea\label{alphaoverlapd}
{\alpha^{(n)}_d}=\int_{SU(d)} dU\left({U\ket{e}\bra{e} U^\dagger}\right)^{\otimes n+1} \!\!\nn\\\otimes \left({U U_0\ket{e}\bra{e} U_0^\dagger U^{\dagger}}\right)^{\otimes n}
\eea\label{betaoverlapd}
\bea{\beta^{(n)}_d}=\int_{SU(d)} dU\left({U\ket{e}\bra{e} U^\dagger}\right)^{\otimes n} \!\! \nn\\\otimes \left({U U_0\ket{e}\bra{e} U_0^\dagger U^{\dagger}}\right)^{\otimes n+1}.
\eea
By the invariance of the Haar measure, one can insert for free an integration over an $SU(2)$ subgroup of $SU(d)$ which act non-trivially only on the space $\mathcal E$ generated by $\ket{e}$ and $U_0\ket{e}$, and write
\begin{multline}\label{alphaoverlapd}
{\alpha^{(n)}_d}=\int_{SU(d)} dU \int_{SU(2)}dV\left({UV\ket{e}\bra{e} V^\dagger U^\dagger}\right)^{\otimes n+1} \!\!\\\otimes \left({U VU_0\ket{e}\bra{e} U_0^\dagger V^\dagger U^{\dagger}}\right)^{\otimes n}=\\
=\int_{SU(d)} dU U^{\otimes 2 n+1}\alpha^{(n)}_2{U^\dagger}^ {\otimes 2 n+1},
\end{multline}

\begin{multline}\label{betaoverlapd}
{\beta^{(n)}_d}=\int_{SU(d)} dU \int_{SU(2}dV\left({UV\ket{e}\bra{e} V^\dagger U^\dagger}\right)^{\otimes n} \!\!\\\otimes \left({UV U_0\ket{e}\bra{e} U_0^\dagger V^\dagger U^{\dagger}}\right)^{\otimes n+1}=\\
=\int_{SU(d)} dU U^{\otimes 2 n+1}\beta^{(n)}_2{U^\dagger}^ {\otimes 2 n+1},
\end{multline}

with $\alpha^{(n)}_2$ and $\beta^{(n)}_2$ defined as in (\ref{alphaoverlap}), (\ref{betaoverlap}), but supported on $\mathcal E ^{\otimes 2n+1}$. By the same token, one has

\begin{multline}\label{thetad}
\Theta^{}_d=\alpha^{(n)}_d-\beta^{(n)}_d= \\\int_{SU(d)} dU U^{\otimes 2 n+1}(\alpha^{(n)}_2-\beta^{(n)}_2){U^\dagger}^ {\otimes 2 n+1}=\\\int_{SU(d)} dU U^{\otimes 2 n+1}\Theta^{}_2{U^\dagger}^ {\otimes 2 n+1}.
\end{multline}

Since Schur-Weyl duality holds also for $SU(d)$, the Hilbert space of $2n+1$ multiple systems still admits a decomposition 

\be\label{ShWdual}
\mathcal H= \bigoplus_{D} (\gamma_{D} \otimes \mu_{D})
\ee
  where $\gamma_{D}$ and $\mu_{D}$ are the irreducible representations of $SU(d)$ and the symmetric group $S_{2n+1}$ with Young diagram $D$.
  
Nonetheless, $\mathcal E^{\otimes 2n+1}$ can be decomposed as $\mathcal E^{\otimes 2n+1} = \bigoplus_D j_{D}\otimes \mu_{D}$ with $j_{D}$,  $\mu_{D}$ irreducible representations of $SU(2)$ and $S_{2n+1}$ with Young diagram $D$. In particular it follows that $j_D \subseteq \gamma_D$. From (\ref{overlaptheta}) we know that $\Theta^{}_2$ has the form
\be
\Theta^{}_2=\sum_q \left(\lambda_{q}(+)\Pi_{q,+}^{(\mathcal E)}+\lambda_{q}(-)\Pi_{q,-}^{(\mathcal E)}\right),
\ee
with $\Pi_{q,\pm}^{(\mathcal E)}$ projecting on vectors of the form $\ket{q,m}\otimes\ket{\pm}$, with
$\ket{q,m}\in j_{q}, \ket{\pm} \in \mu_q $, with $q$ being the associated Young diagram.

Therefore, calling $\Pi_{q,+}$ the projector on $\gamma_{q}\otimes\ket{+}\bra{+}$ and $\Pi_{q,-}$ the projector on $\gamma_{q}\otimes\ket{-}\bra{-}$, we have
\begin{multline}
\Theta^{}_d=\int_{SU(d)} dU U^{\otimes 2 n+1}\Theta^{}_2{U^\dagger}^ {\otimes 2 n+1}\\
=\sum_{q}\lambda_{q }(+)\int_{SU(d)} dU {U}^{\otimes{2n+1}}\Pi^{(\mathcal E)}_{q,+}U^{\otimes {2n+1}}\\
+\sum_{q}\lambda_{q }(-)\int_{SU(d)} dU {U}^{\otimes{2n+1}}\Pi^{(\mathcal E)}_{q,-}U^{\otimes {2n+1}}\\
=\sum_{q}\frac{2q+1}{g_q^{(d)}}\left(\lambda_{q }(+)\Pi_{q,+}+\lambda_{q }(-)\Pi_{q,-}\right)\,,\label{rhoc}
\end{multline}

where we used the Peter-Weyl theorem \cite{Knapp} in the last equality, where now $g_q^{(d)}$ is the dimension of the representation $\gamma_q$. Finally, since $\Tr[\Pi_{q,\pm}]=g_q^{(d)}$, the probability of error in the $d$ dimensional case is still (\ref{resultoverlap}):
\bea
 P_{err,min, d}^{(n\gg1)}  \simeq\frac 1 2\left(1-|\cos\tfrac \theta 2|\right)+\tfrac{3 + \cos\theta}{8 \sqrt 2 \sqrt{1 + \cos\theta}}\frac{1}{n}\nn\\+\tfrac{1 -60 \cos\theta-5\cos2\theta}{128 \sqrt 2 (1 + \cos\theta)^{3/2}}\frac{1}{n^2} \;, &&
\eea 
with $\sin\frac \theta 2=|\bra{e}U_0\ket{e}|$. Also in this case one can get the probability of error of the optimal learning machines for Haar random pure states by integrating over the probability distribution of the overlap $c=\sin^2 \frac \theta 2$, which for Haar random $\ket\psi_1$ and $\ket{\psi_2}$ is known (e.g. $\cite{OverlapStatistics}$) and equal to $P(c)=(d-1)(1-c)^{d-2}$. At the next to leading order the result is

\bea
 P_{err,min, d}^{(n\gg1)}  \simeq\frac 1 2-\frac{d-1}{2d-1}+\frac{(d-1)^2}{3+4d(d-2)}\frac 1 n \;. &&.
\eea 

which agrees with the zeroth order result in \cite{HHH1}. The asymptotic correction that we find can be also directly calculated by following the approach in \cite{HHH1}, the interested reader can find the calculations in the supplementary Mathematica notebooks.}

\subsection{Compatibility between optimal machines}\label{SUBSECTIOND} 
In the previous subsections we have analysed three different scenarios, which in principles give rise to different optimal machines. However, additional symmetries make some of the optimal machines compatible, in the sense that it exists a measurement that is optimal for different scenarios. In particular, if $\hat S_{AB}$ is the swap operator between $\mathcal H_A$ and $\mathcal H_B$, one can verify that $\hat S_{AB}\alpha^{(n)}\hat S_{AB}^\dagger=\beta^{(n)}$ in scenario (ii), (iii) and also (i) when $r_1=r_2$. If this happens then $\hat S_{AB} \Theta^{}\hat S_{AB}^\dagger=-\Theta^{}$; it follows that if $\ket{\lambda}$ is an eigenvector of $\Theta^{}$ with eigenvalue $\lambda$, then also $\hat S_{AB}\ket{\lambda}$ is an eigenvector, with eigenvalue $-\lambda$. Since $\hat S_{AB}\ket{s+1/2,s;q,m}_{i,k}=\ket{s,s+1/2;q,m}_{i,k}$, $\hat S_{AB}\ket{s-1/2,s;q,m}_{i,k}=\ket{s,s-1/2;q,m}_{i,k}$, in the spaces ${\cal H}_{A_iXB_k}^{(s,s)}$ the eigenvectors are automatically determined as the orthogonal vectors $\ket{\lambda_+}$,$\ket{\lambda_-}$ in ${\cal H}_{A_iXB_k}^{(s,s)}$ such as $\hat S_{AB}\ket{\lambda_+}=\ket{\lambda_-}$.

In particular, since the the relevant subspace in scenario (iii) is only ${\cal H}_{A_iXB_k}^{(\frac n 2,\frac n 2 )}$, the optimal machine for scenario (i) when $r_1=r_2$, or the one for scenario (ii), are also optimal for scenario (iii).

\section{Implementation of the optimal POVM} \label{SEC:povm}
From the knowledge of the eigenvectors (\ref{eigvector}) one can reconstruct the optimal POVM. Since it is a projective measurement, it can be realized by a change of basis from the the eigenvectors to the computational basis, followed by a local measurement. In the following we consider the implementation of the optimal machine of scenario iii), for the case $n=1$. The change of basis is:
\bea
|\psi_{\tfrac 1 2,\tfrac 1 2,;\tfrac 3 2,\tfrac 3 2}\rangle\rightarrow\ket{\uparrow\uparrow\uparrow} \,\,\,\,(C)\nn\\
|\psi_{\tfrac 1 2,\tfrac 1 2,;\tfrac 3 2,\tfrac 1 2}\rangle\rightarrow\ket{\uparrow\uparrow\downarrow}\,\,\,\,(C)\nn\\
|\psi_{\tfrac 1 2,\tfrac 1 2,;\tfrac 1 2,\tfrac 1 2}^{(-)}\rangle\rightarrow\ket{\downarrow\uparrow\uparrow}\,\,\,\,(B)\nn\\
|\psi_{\tfrac 1 2,\tfrac 1 2,;\tfrac 1 2,\tfrac 1 2}^{(+)}\rangle\rightarrow\ket{\uparrow\downarrow\uparrow}\,\,\,\,(A)\nn\\
|\psi_{\tfrac 1 2,\tfrac 1 2,;\tfrac 1 2,-\tfrac 1 2}^{(+)}\rangle\rightarrow\ket{\downarrow\uparrow\downarrow}\,\,\,\,(A)\nn\\
|\psi_{\tfrac 1 2,\tfrac 1 2,;\tfrac 1 2,-\tfrac 1 2}^{(-)}\rangle\rightarrow\ket{\uparrow\downarrow\downarrow}\,\,\,\,(B)\nn\\
|\psi_{\tfrac 1 2,\tfrac 1 2,;\tfrac 3 2,-\tfrac 1 2}\rangle\rightarrow\ket{\downarrow\downarrow\uparrow}\,\,\,\,(C)\nn\\
|\psi_{\tfrac 1 2,\tfrac 1 2,;\tfrac 3 2,-\tfrac 3 2}\rangle\rightarrow\ket{\downarrow\downarrow\downarrow}\,\,\,\,(C)\nn\\
\eea

where $A$ ($B$) means that the result of the measurement is interpreted as $X=A$ ($X=B$), while for $C$ we "flip a coin" to decide. In the computational basis the unitary rotation reads
\bea\left(
\begin{smallmatrix}
1& 0& 0& 0& 0& 0& 0& 0\\ 0& \frac{1}{\sqrt{3}}& \frac{1}{\sqrt{3}}& 0& \frac{1}{\sqrt{3}}& 0& 
  0& 0\\ 0& \frac{1}{\sqrt{3}}& \frac{-3 - \sqrt{3}}{6}& 0&  \frac{3 - \sqrt{3}}{6}& 0& 
  0& 0\\0& 0& 0& \frac{-3 - \sqrt{3}}{6}& 0& \frac 1{3 + \sqrt{3}}& \frac{1}{\sqrt{3}}& 
  0\\0& \frac{1}{\sqrt{3}}& \frac{3 - \sqrt{3}}{6}& 0& \frac{-3 -\sqrt{3}}{6}& 
  0& 0& 0\\0& 0& 0& \frac{3 - \sqrt{3}}{6}& 0& \frac{1}{-3 + \sqrt{3}}& \frac 1 {\sqrt{
  3}}& 0\\0& 0& 0& \frac{1}{\sqrt{3}}& 0& \frac{1}{\sqrt{3}}& \frac{1}{\sqrt{3}}& 0\\0& 0& 0& 
  0& 0& 0& 0& 1
\end{smallmatrix}
\right),
\eea
and the probability of error as a function of $\theta$ is

\be
P_{err,min}^{(1)}=\frac 1 2 -\frac{1+\cos\theta}{4\sqrt{3}}.
\ee

These kind of operations are suitable for all programmable devices which are based on the circuit model of quantum computation, as for example the recent quantum chips developed by IBM \cite{qchip}.  By using the software development kit QISKit \cite{qiskit}, we have determined a circuit that realises the POVM for the $n=1$ case with input pure states and checked its performance with the IBM simulator. The number of gates of our implementation is $61$ single qubit operations and $60$ CNOT, with a depth of $43$ operations. Given that the failure probability of a CNOT on real machines is about $5*10^{-2}$, the failure probability of the circuit is at least $1-0.95^{60}\approx 0.954$. Indeed we tried to remotely perform the experiment on the real physical chip, without any significant results. This fact underlines the importance of gate optimisation and error correction for the proper operation of future quantum computers.
However, with the simulation tools of QISKit Aer, we were able to simulate the circuit with an error model consisting in depolarising errors (Fig.~\ref{gate_error}) and  thermal relaxation errors (Fig.~\ref{therm_relax}): decreasing the depolarising probability and increasing the relaxation times we can show how the circuit is sensitive to this kind of noises, and that we recover the expected behaviour for small noise.

\begin{figure}
    \centering
        \includegraphics[width=0.47\textwidth]{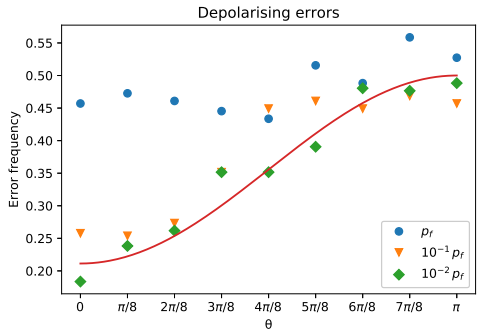}
\caption{\label{gate_error} {\small{\emph{Simulation of the optimal machine with QISKit Aer, with depolarising error modeled after the gate average infidelity of each gate: $p_f$ are the depolarising probability for the 16 qubit machine (Melbourne) if all the infidelity is due to a depolarising channel. Frequency of misclassification errors with 256 repetitions for each $\theta$, compared with the predicted minimum error function (solid line). } }} }
\end{figure}

\begin{figure}
    \centering
        \includegraphics[width=0.47\textwidth]{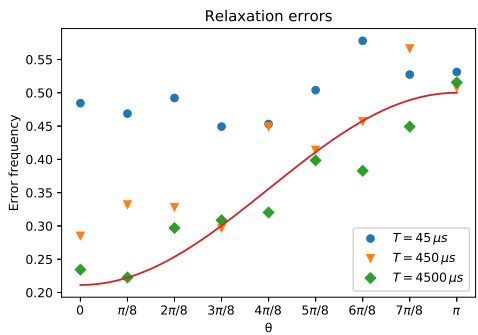}
\caption{\label{therm_relax} {\small{\emph{Simulation of the optimal machine with QISKit Aer, with thermal relaxation times $T_1=T_2=T$ equal for each qubit. Gate times are set to $200\, ns$ for 1 qubit gates and $800\, ns$ for CNOT. Frequency of misclassification errors with 256 repetitions for each $\theta$, compared with the predicted minimum error function (solid line).  } }} }
\end{figure}

\section{Conclusions} \label{SEC:con} 

In this work we have discussed the performances  of optimal universal learning quantum machines that aim at discriminating the states of a qudit
starting from a collection of templates states in the hybrid, yet realistic scenario, where at least some global information on the training set is classically available. 
As a matter of fact, it is not hard to identify situations for which  this kind of approach could provide a realistic modelisation. Indeed, while absolute information about quantum states
is typically not accessible, some structural properties are more likely to be available. For instance this is what happens in quantum communication \cite{HOLEVOBOOK} where
the receiving party does not know the particular state is going to receive, but has classical knowledge on the code the sender is using. 
Like classical supervised learning is a fundamental tool with classical data, arguably quantum learning machines will be important for dealing with quantum data with quantum processors. Indeed, given that quantum tomography is very expensive in terms of resources, dealing with quantum data requires to study alternatives which need little information about the data, make use of the full power of quantum mechanics, and extract only the relevant information for the problem at hand. Our work extends the previous results considering more general scenarios. 
An interesting observation is that the optimal machine that does not assume any kind of information about the template state, scenario (ii), it is also optimal for scenario (iii), where the template states are assumed to be pure. It is therefore a very general machine, which can be seen as the most convenient learning algorithm.

\section*{Acknowledgements}

The Authors thank Gael Sent\'{i}s for a careful reading of the manuscript and fruitful comments.

\section{Appendix}\label{appHaar}

In this appendix we present the explicit derivation of some important identities which are used in the main text. 
We recall that the Haar measure $dU$ of $SU(2)$  satisfies the identity
$$\int dU \,{\cal F}(LU)=\int dU\, {\cal F}(U) =\int dU\, {\cal F}(UR)\;,$$
for all  $L,R\in SU(2)$ and for all  functions $\cal F$ acting on $SU(2)$, 
and that it induces a Hilbert product on $L_2[U,dU]$ via the identification
$$({f},{g})\equiv \int dU \, f^*(U) g(U)\;.$$
Furthemore, indicating with $D^j_{m,m'} (U)$  the matrix elements of the irreducible representations of $U\in SU(2)$ with Casimir $j(j+1)$,  we recall that
via the 
Peter-Weyl theorem they fulfil the identities 
\bea&& \int dU\, {\left(D^{j_1}_{m_1,m_1'}(U)\right)}^*D^{j_2}_{m_2,m_2'}(U) \label{PETER}  \\ \nn 
&&\qquad \qquad \qquad =\frac 1 {2j_1+1} \delta_{j_1 j_2} \delta_{m_1 m_2}\delta_{m_1'm_2'}\;.\eea

\subsection{Derivation of Eq.~(\ref{operatoraverage})} \label{APPA1} 

Let $\rho$ a qubit density matrix characterized by  Bloch vector of length $r$ which, without loss of generality we shall assume to be  oriented in the positive $\hat{z}$ direction, i.e. 
$\rho=\left(\frac{1+r}{2} \right) \ket \uparrow\bra \uparrow+\left(\frac{1-r}{2}\right)  \ket \downarrow\bra \downarrow$ with  
$\ket \uparrow$, $\ket \downarrow$ being the eigenvectors of $\sigma_z$. 
We notice that its $n$-th tensor power can be expressed as
\begin{eqnarray} 
\rho^{\otimes n} = \sum_{l=0}^{n} \left(\frac{1+r}{2}\right)^{l} \left(\frac{1-r}{2}\right)^{n-l} B^{(n)}_l,\nn 
\end{eqnarray} 
with 
\bea
 B^{(n)}_l \equiv \sum_{\pi}S_\pi\left( \ket \uparrow\bra \uparrow^{\otimes l}\otimes \ket \downarrow\bra \downarrow^{\otimes n-l} \right)S^\dagger_\pi\;, \nn 
\eea 
the sum being performed over the set of permutations operators $S_\pi$  of $n$ elements.
By construction $B^{(n)}_l$ is the projector on the eigenspace at fixed total angular momentum $J_z$, therefore it is diagonal in every basis of eigenvectors of $J^2,J_z$.
In particular its support is given by the vectors $\ket{j,l-\frac n 2}_i$ in each representation with Casimir number $J^2=j(j+1)$ and $l\in\{\frac n 2 -j, \cdots, \frac n 2 +j\}$, the index
  $i$ labelling accounting for the multiplicity  of the representation, i.e. 
\bea
 B^{(n)}_l = \oplus_{j\geq |l-\tfrac n 2|} \oplus_i \ket{j,l-\tfrac n 2}_i\bra{ j,l- \tfrac n 2}\;.
\eea 
Consider then the operator
\bea 
\gamma^{(n)}&\equiv&\int dU \left(U\rho U^\dagger\right)^{\otimes n}  \\
&=&\sum_{l=0}^{n} \left(\frac{1+r}{2}\right)^{l} \left(\frac{1-r}{2}\right)^{n-l} P^{(n)}_l,\nn 
\eea
with  
\bea P^{(n)}_l &\equiv&  \int dU U^{\otimes n} B^{(n)}_l{U^\dagger}^{\otimes n}\;.
\eea 
Invoking  the identity~(\ref{PETER}) we can conclude that 
\bea
P^{(n)}_l&=&\oplus_{j\geq |l-\tfrac n 2|} \frac {\mathbf 1^{(j)}} {2 j+1},
\eea
where now $\mathbf 1^{(j)}$ is the projector on all irreducible representation with principal quantum number $j$.
Accordingly we have  
\bea\label{FF11221} 
\gamma^{(n)}&=&\oplus_{j} {\sum_{l=\tfrac n 2-j}^{\tfrac n 2+j} \left(\frac{1+r}{2}\right)^{l} \left(\frac{1-r}{2}\right)^{n-l}} \frac{\mathbf 1^{(j)}}{2j+1} \nn\\&=&\oplus_{j}f_j^{(n)}(r)\mathbf 1^{(j)},
\eea
with $f_j^{(n)}(r)$ as in (\ref{DEFFUNZIONE}). 
Equation~(\ref{operatoraverage}) finally follows from by a direct application of  (\ref{FF11221}) to the terms 
$\int dU \left(U\rho_1 U^\dagger\right)^{\otimes n+1}$ and $\int dU \left(U\rho_2 U^\dagger\right)^{\otimes n}$ that enter in the definition of the operator $\alpha^{(n)}$ of 
Sec.~\ref{SEC:i}. 

\subsection{Derivation of Eq.~(\ref{fixangle})} \label{APPAB}

To derive  (\ref{fixangle}) let us first expand  $\ket{\tfrac{n+1}2, \tfrac{n}2; q,m}$ into the angular momentum basis 
given by the tensor product states $\ket{\frac{n+1}{2},m'}\otimes \ket{\frac{n}{2},m-m'}$ associated with the $AX/B$ partition, i.e. 
\begin{eqnarray}  \label{CLEB}
\ket{\tfrac{n+1}2, \tfrac{n}2; q,m} &=& \sum_{m'} C^{q,m}_{\frac{n+1}{2},m',\frac{n}{2},m-m'}\\
&& \times \ket{\frac{n+1}{2},m'}\otimes \ket{\frac{n}{2},m-m'} \;, \nn \end{eqnarray} 
where $C^{q,m}_{\frac{n+1}{2},m',\frac{n}{2},m-m'}$ are the corresponding Clebsch-Gordan coefficients. 
Then observing that in this basis the state $\ket{\uparrow}^{\otimes n+1} \otimes \ket{\uparrow}^{\otimes n}$ corresponds to the element
 $\ket{\tfrac{n+1}{2},\tfrac{n+1}{2}} \otimes \ket{\tfrac{n}{2},\tfrac{n}{2}}$, we write 
the operator $\alpha^{(n)}$  as 
\begin{eqnarray} {\alpha^{(n)}}=\int dU U^{\otimes n+1} \nn
\ket{\tfrac{n+1}{2},\tfrac{n+1}{2}}\bra{\tfrac{n+1}{2},\tfrac{n+1}{2}} {U^\dagger}^{\otimes n+1} 
\\
\otimes (U U_0)^{\otimes n}\nn
\ket{\tfrac{n}{2},\tfrac{n}{2}}\bra{\tfrac{n}{2},\tfrac{n}{2}} ({U_0^\dagger U^\dagger})^{\otimes n} \;,\end{eqnarray} 
and observe that 
\bea
&&\!\!\!{\alpha^{(n)}} \left( \ket{\tfrac{n+1}{2},m'}\otimes \ket{\tfrac{n}{2},m-m'}\right)= \nn\\
&=& \!\!\!\sum_{l,l'}  \int dUD^{\frac{n+1}{2}}_{l, \frac{n+1}{2} }(U){D^{\frac{n+1}{2}}_{\frac{n+1}{2},m'}}(U^\dagger)  \nn\\
&\times& D^{\frac{n}{2}}_{l' ,\frac{n}{2}}(UU_0) {D^{\frac{n}{2}}_{\frac{n}{2},m-m'}}(U_0^\dagger U^\dagger)  \ket{\tfrac{n+1}{2},l}\otimes \ket{\tfrac{n}{2},l'}\nn\\
&=&\!\!\!\sum_{h,k} D^{\frac{n}{2}}_{h, \frac{n}{2}}(U_0) {D^{\frac{n}{2}}_{\frac{n}{2},k}}(U_0^\dagger)\nn\\&\times& \sum_{l,l'}  \int dU D^{\frac{n+1}{2}}_{l ,\frac{n+1}{2} }(U){D^{\frac{n+1}{2}}_{\frac{n+1}{2},m'}}(U^\dagger)
\nn \\ 
&\times&  D^{\frac{n}{2}}_{l' , h}(U) {D^{\frac{n}{2}}_{k ,m-m'}}(U^\dagger)  \ket{\tfrac{n+1}{2},l}\otimes \ket{\tfrac{n}{2},l'} \;, \eea
where in the first identity the matrix elements 
$$D^{\frac{n+1}{2}}_{m_1, m_2}(U)=\bra{\frac{n+1}{2},m_1} U^{\otimes {n+1}}\ket{\frac{n+1}{2},m_2},$$  
and 
$$D^{\frac{n}{2}}_{m_1, m_2}(U)=\bra{\frac{n}{2},m_1} U^{\otimes {n}}\ket{\frac{n}{2},m_2},$$  
represent the action of the unitary $U$ 
  into the selected basis,  
while in the second we used the composition rules of $SU(2)$ to factorize the contributions of $U_0$ from the rest. 
This equation can be further simplified by exploiting once more the Clebsch-Gordan mapping (\ref{CLEB}) to 
merge together $D^{\frac{n+1}{2}}_{l ,\frac{n+1}{2} }(U)$ with $D^{\frac{n}{2}}_{l' , h}(U)$, and ${D^{\frac{n+1}{2}}_{\frac{n+1}{2},m'}}(U^\dagger)$ with ${D^{\frac{n}{2}}_{k ,m-m'}}(U^\dagger)$.
As a result the previous expression becomes 
\bea
&&\!\!\!{\alpha^{(n)}} \left( \ket{\tfrac{n+1}{2},m'}\otimes \ket{\tfrac{n}{2},m-m'}\right) \nn \\
&&= \sum_{h,k} D^{\frac{n}{2}}_{h ,\frac{n}{2}}(U_0) {D^{\frac{n}{2}}_{\frac{n}{2},k}}(U_0^\dagger) \nn \\ &&\times 
\sum_{l,l',q,q'}  \int dU D^{q}_{l+l',\frac{n+1}{2}+h}(U)D^{q'}_{\frac{n+1}{2}+k,{m}}(U^\dagger)\nn
\\&&
\times   C^{{q},l+l'}_{\frac{n+1}{2},l,\frac{n}{2},l'}C^{{q},\frac{n+1}{2}+h}_{\frac{n+1}{2},\frac{n+1}{2},\frac{n}{2},h} \nn
\\ &&
 \times C^{{q'},{m}}_{\frac{n+1}{2},m',\frac{n}{2},{m}-m'}  C^{{q'},\frac{n+1}{2}+k}_{\frac{n+1}{2},\frac{n+1}{2},\frac{n}{2},k} \ket{\tfrac{n+1}{2},l}\otimes \ket{\tfrac{n}{2},l'}\nn\\
&&=\sum_{l,q}   \frac {1}{2q+1} \sum_{h} D^{\frac{n}{2}}_{h ,\frac{n}{2}}(U_0) {D^{\frac{n}{2}}_{\frac{n}{2},h}}(U_0^\dagger )\nn \\ &&\times 
 C^{{q},{m}}_{\frac{n+1}{2},m',\frac{n}{2},{m}-m'} C^{{q},\frac{n+1}{2}+h}_{\frac{n+1}{2},\frac{n+1}{2},\frac{n}{2},h} \nn
\\ &&
 \times C^{{q},\frac{n+1}{2}+h}_{\frac{n+1}{2},\frac{n+1}{2},\frac{n}{2},h}C^{{q},{m}}_{\frac{n+1}{2},l,\frac{n}{2},m-l}  \ket{\tfrac{n+1}{2},l}\otimes \ket{\tfrac{n}{2},m-l}\nn\;,
\eea
where in the second identity we exploit the Peter-Weyl theorem, see Eq.~(\ref{PETER}), to evaluate the integral in $U$, obtaining that the nonzero terms in the sum have to satisfy $q=q'$, $l+l'=m$, $h=k$.
Multiplying this by $C^{q,m}_{\frac{n+1}{2},m',\frac{n}{2},m-m'}$ while summing over $m'$, the latter equation finally gives us 
\bea 
&&\!\!\!{\alpha^{(n)}} \ket{\tfrac{n+1}2, \tfrac{n}2; q,m} \nn\\ 
&&=  \sum_{l,m',q'} \frac {1}{2q'+1} \sum_{h} D^{\frac{n}{2}}_{h ,\frac{n}{2}}(U_0) {D^{\frac{n}{2}}_{\frac{n}{2},h}}(U_0^\dagger) \nn \\ &&\times 
 C^{q,m}_{\frac{n+1}{2},m',\frac{n}{2},m-m'} C^{{q'},{m}}_{\frac{n+1}{2},m',\frac{n}{2},{m}-m'}  \nn \\ &&\times\, C^{{q'},\frac{n+1}{2}+h}_{\frac{n+1}{2},\frac{n}{2},\frac{n}{2},h} C^{{q'},\frac{n+1}{2}+h}_{\frac{n+1}{2},\frac{n+1}{2},\frac{n}{2},h} \nn \\ &&\times \,C^{{q'},{m}}_{\frac{n+1}{2},l,\frac{n}{2},m-l}  \ket{\tfrac{n+1}{2},l}\otimes \ket{\tfrac{n}{2},m-l}\nn\;, \\
&&=\frac {1}{2{q}+1} \sum_{h}  D^{\frac{n}{2}}_{h ,\frac{n}{2}}(U_0) {D^{\frac{n}{2}}_{\frac{n}{2},h}}(U_0^\dagger)\nn\\&& \times \,C^{{q},\frac {n+1} 2+h}_{\frac{n+1}{2},\frac{n+1}{2},\frac{n}{2},h} C^{{q},\frac {n+1} 2+h}_{\frac{n+1}{2},\frac{n+1}{2},\frac{n}{2},h} \ket{\tfrac{n+1}2, \tfrac{n}2; q,m} \;, \nn\\
\eea
which coincides with (\ref{fixangle}). Here in the second equality we used the orthogonality of the Clebsch-Gordan coefficients to select only the term $q'=q$ in the sum.

\subsection{Central momenta of the distributions } \label{APPAC} 
Here we report  the central momenta of the distributions used for computing (\ref{fixresult}).

Putting $h=\frac{ns} 2$ and  $\frac{1+r}{2}=\cos\left(\frac {\pi-\theta}{2}\right)$ we notice that 
the distribution
\bea
P_\theta^{(n)}(s)&\equiv&D^{\frac{n}{2}}_{h, \frac{n}{2}}(U_0)D^{\frac{n}{2}}_{\frac{n}{2}, h}(U_0^\dagger)\;,\eea
defined in Eq.~(\ref{BINOMIAL}) has momenta 
\bea 
\mu_1&=&E[s]=r\;,\nn\\
\mu_2&=&E[(s-\mu_1)^2]=\frac{1-r^2}{n}\;,\nn\\
\mu_3&=&E[(s-\mu_1)^3]=2r\frac{1-r^2}{n}\;,\nn\\
\mu_4&=&E[(s-\mu_1)^4]\nn\\&=& \frac{(-1 + r^2) (2 - 6 r^2 + 3 n (-1 + r^2))}{n^3}\;.\nn
\eea
Instead, setting $h=\frac {n s}{2}$, we notice that  the momenta of the distribution
\bea
&P^{(n)}_h(q)\equiv\frac{2 \left(\frac n 2 -h\right)!\left(n+1\right)!}{
\left(\frac n 2 +h\right)!}&\nn\\&\times\frac{\left(\frac n 2 +h+q+\frac 1 2\right)!}{
\left(q-\frac 1 2-\frac n 2 -h\right)!\left(n-q+\frac 1 2\right)!\left(n+q+\frac 3 2\right)!}\;,& \nn\\
\eea
defined in~(\ref{NEWDIST}), can be expressed in terms of Euler gamma functions as follows 
\bea
\mu_1&=&E[q]\nn\\&=&-\frac 1 2 + \frac{\Gamma(1/2 + h + n/2) \Gamma(2 + n)}{
 \Gamma(1 + h + n/2) \Gamma(3/2 + n)} \nn\\
 &=& \frac{n \sqrt{1 + s}}{\sqrt 2} -\frac 1 2  \nn\\ &+& \frac{11 + 5 s}{
 8 \sqrt{2} \sqrt{1 + s}} + \frac{9 + 14 s - 23 s^2}{
 128 \sqrt{2} n (1 + s)^{3/2}}\nn\\&+&O\left(\frac 1 {n^2}\right)\;,\nn\\
 \eea
 \bea
\mu_2&=&E[(q-\mu_1)^2]\nn\\ &=&\frac 1 2 (1 + n) (2 + 2 h + n) \nn\\ &-& \frac{\Gamma(3/2 + h + n/2)^2 \Gamma(2 + n)^2}{
 \Gamma(1 + h + n/2)^2 \Gamma(3/2 + n)^2}\nn\\
 &=&\frac 1 8 n (1 - s) + \frac{-1 + 2 s - s^2}{64 (1 + s)}+O\left(\frac 1 {n}\right)\nn\\
  \eea
 \bea
& \mu_3=E[(q-\mu_1)^3]\nn\\ &=\frac{-(8 + 2 h (5 + 4 n) + n (11 + 4 n))\Gamma(
     3/2 + h + n/2)\Gamma(2 + n)}{
 \Gamma(1 + h + n/2) \Gamma(3/2 + n)} &\nn\\
 &+ 
 \frac{  8 \Gamma(3/2 + h + n/2)^3 \Gamma(2 + n)^3}{4 \Gamma(
    1 + h + n/2)^3 \Gamma(3/2 + n)^3}\nn\\
     &=\frac{(-1 + s)^2 n}{32 \sqrt 2  \sqrt{1 + s}}+O\left(1\right)\;,&\nn\\
      \eea
 \bea
&\mu_4=E[(q-\mu_1)^4]\nn\\ &= \frac {(1 + n) (4 + 10 n + 4 h^2 n + 6 n^2 + n^3 + 
    4 h (1 + 3 n + n^2))}{4}&\nn\\& - \frac{3 \Gamma(3/2 + h + n/2)^4 \Gamma(2 + n)^4}{
 \Gamma(1 + h + n/2)^4 \Gamma(3/2 + n)^4} &\nn\\&+ \frac{
(2 + 2 n + n^2 + 2 h (2 + n)) \pi^2 \Gamma(
   2 + 2 h + n)^2 \Gamma(3 + 2 n)^2}{ 4^{3 + 2 h + 3 n} 
 \Gamma(1 + h + n/2)^4 \Gamma(3/2 + n)^4}&\nn\\
  &=\frac 3 64 (1 - 2 s + s^2) n^2 +O\left(n\right)\;.&\nn
  \\
\eea

\end{document}